\documentclass[preprint,showpacs,preprintnumbers,amsmath,amssymb]{revtex4}

\usepackage{graphicx}
\usepackage{dcolumn}
\usepackage{bm}
\usepackage{color}

\usepackage{feynmf}
\usepackage{slashed}

\usepackage[utf8]{inputenc}

\begin{document}

\title{The $\nu$THDM with the Inverse Seesaw Mechanisms}

\author{Yi-Lei Tang}
\thanks{tangyilei15@pku.edu.cn}
\affiliation{Center for High Energy Physics, Peking University, Beijing 100871, China}


\author{Shou-hua Zhu}
\thanks{shzhu@pku.edu.cn}
\affiliation{Institute of Theoretical Physics $\&$ State Key Laboratory of Nuclear Physics and Technology, Peking University, Beijing 100871, China}
\affiliation{Collaborative Innovation Center of Quantum Matter, Beijing 100871, China}
\affiliation{Center for High Energy Physics, Peking University, Beijing 100871, China}

\date{\today}

\begin{abstract}

In this paper, we combine the $\nu$-Two-Higgs-Doublet-Model ($\nu$THDM) with the inverse seesaw mechanisms. In this model, the Yukawa couplings involving the sterile neutrinos and the exotic Higgs bosons can be of order one in the case of a large $\tan \beta$. We calculated the corrections to the Z-resonance parameters $R_{l_i}$, $A_{l_i}$, $N_{\nu}$, together with the $l_1 \rightarrow l_2 \gamma$ branching ratios, and the muon anomalous $g-2$. Compared with the current bounds and plans for the future colliders, we find that the corrections to the electroweak parameters can be contrained or discovered in much of the parameter space.

\end{abstract}


\maketitle
\section{Introduction}

The smallness of the neutrino masses can be explained by the seesaw mechanisms. In the framework of the Type-I seesaw mechanisms \cite{SeeSaw1, SeeSaw2, SeeSaw3, SeeSaw4, SeeSaw5}, large Majorana masses ($\sim M_N$) are introduced for the right-handed neutrinos. The Yukawa couplings ($y_D L H N$) between the left-handed and the right-handed neutrinos through a Higgs doublet generate the Dirac masse terms ($\sim m_D=y_D v$). After ``integrating out'' the right-handed neutrinos, or equivalently diagonalizing the full neutrino mass matrix, one obtain the tiny neutrino masses ($\sim \frac{m_D^2}{M_N}$) suppressed by the $M_N$ in the denominator.

The standard seesaw mechanisms usually require extremely large $M_N \sim 10^{9\text{-}13} \text{ GeV}$ in the case that the Yukawa coupling constant $y_D \sim 0.01\text{-}1$, which is beyond the scope of any realistic collider proposal. An alternative scheme to lower the sterile neutrinos masses down to the 100-1000 GeV scale without introducing too small Yukawa coupling constants is the ``inverse seesaw'' mechanisms \cite{Inverse1, Inverse2, Inverse3, Inverse4}. In the inverse seesaw mechanisms, pairs of the weyl-spinors charged with the lepton number (L) form the pseudo-Dirac neutrinos ($N_{L,R}$). Small majorana mass terms ($\sim \mu \overline{N}_L N_L^c$) which softly break the lepton number are introduced as well as the lepton-number-conserving Dirac mass terms ($\sim m_{N} \overline{N}_L N_R$). Again, after integrating out the sterile neutrinos, or equivalently diagonalizing the full neutrino mass matrix, one find the tiny neutrino masses ($\sim \frac{m_D^2}{m_{N}^2} \mu$). Thus, the smallness of the neutrino masses is explained by the smallness of the $\mu$.

Compared with the standard TeV-scale seesaw mechanisms, the mixings between the left-handed and the sterile neutrinos can be much larger in the inverse seesaw mechanisms. This offers us some possibilities to test or constrain the models by the collider experiments. However, the $LHN_D$ Yukawa couplings should still be well below the order of one due to various constraints. One way to raise the Yukawa coupling constants of the neutrinos is the $\nu$-two-Higgs-doublet model ($\nu$THDM) (For some early works, see Ref.~\cite{vTHDM_Ma, vTHDM_Ancester}. For some discussions of the collider physics, see Ref.~\cite{vTHDM_Collider1, vTHDM_Collider2}. For a variant, see Ref.~\cite{vTHDM_Dirac1, vTHDM_Dirac2}.). This is a variant of the Type-I two-Higgs-doublet model (For a review of the THDM, see \cite{THDM_Review}, and for references therein). In this model, all the standard model fermions couple with one of the Higgs doublet (usually named $\Phi_2$), while the neutrino sector couples with the other ($\Phi_1$). The Yukawa coupling constants of the neutrino sector are then amplified by a factor of $\sec\beta \approx \tan\beta = \frac{v_2}{v_1}$. In the usual cases of the $\nu$THDM, we need a $\tan\beta \gtrsim 10^4$ in order for a Yukawa coupling of order one. However, if we combine the $\nu$THDM with the inverse seesaw mechanisms, a $\tan\beta \sim 10^{2\text{-}3}$ is enough.

The relatively large Yukawa coupling constants will not only provide the opportunities of directly observing the sterile neutrinos in the future collider experiments, but will also show up some electroweak observables. In this paper, we concentrate on the Z-resonance observables $R_l$ and $A_l$, where $l$=e, $\mu$, $\tau$ (Besides the corresponding chapters in the Ref.~\cite{PDG}, see Ref.~\cite{ZResonancePR, ZResonanceOrdinary, Muongm2_1, Muongm2_2, Muongm2_3} for the details). We also consider the leptonic flavor changing neutral current (FCNC) $l_1 \rightarrow l_2 + \gamma$ decay bounds, the muon anomalous magnetic moment, and . We will show that in some of the parameter space, it is possible for the future collider experiments to detect the small deviations on Z-resonance observables originated from this model.

\section{Model Descriptions}

Beforehand, we shall make a brief review of the THDM. The Higgs potential is given by
\begin{eqnarray}
  V&=& m_1^2 \Phi_1^{\dagger} \Phi_1 + m_2^2 \Phi_2^{\dagger} \Phi_2 - m_{12}^2 (\Phi_1^{\dagger} \Phi_2 + \Phi_2^{\dagger} \Phi_1) + \frac{\lambda_1}{2} (\Phi_1^{\dagger} \Phi_1)^2 + \frac{\lambda_2}{2} (\Phi_2^{\dagger} \Phi_2)^2 \nonumber \\
  &+& \lambda_3 (\Phi_1^{\dagger} \Phi_1)(\Phi_2^{\dagger} \Phi_2) + \lambda_4 (\Phi_1^{\dagger} \Phi_2)(\Phi_2^{\dagger} \Phi_1) + \frac{\lambda_5}{2} \left[ (\Phi_1^{\dagger} \Phi_2)^2 + (\Phi_2^{\dagger} \Phi_1)^2 \right],
\end{eqnarray}
where $\Phi_{1,2}$ are the two Higgs doublets with hypercharge $Y=\frac{1}{2}$, $\lambda_{1\text{-}7}$ are the coupling constants, $m_1^2$, $m_2^2$ and $m_{12}^2$ are the mass parameters. As in most of the cases in the literature, we impose a $Z_2$ symmetry that $\Phi_i \rightarrow (-1)^{i-1} \Phi_i$ to avoid the tree-level FCNC. This symmetry forbid the $\left[\lambda_6 (\Phi_1^{\dagger} \Phi_1) + \lambda_7 (\Phi_2^{\dagger} \Phi_2)\right](\Phi_1^{\dagger} \Phi_2+\text{h.c.})$ terms and is softly broken by the $m_{12}^2$ term.

After the electroweak symmetry breaking, the Higgs doublets acquire the vacuum expectation values (VEVs) $v_{1,2}$, and the Higgs component fields form physical mass eigenstates $H^{\pm}$, $h$, $H$, $A$, as well as the Goldstone bosons $G^{\pm, 0}$.
\begin{eqnarray}
\Phi_1 &=& \frac{1}{\sqrt{2}} \left(
\begin{array}{c}
\sqrt{2} (G^{+} \cos\beta -H^{+} \sin\beta) \\
v\cos\beta - h\sin\alpha + H\cos\alpha + i(G^0 \cos\beta - A\sin\beta)
\end{array}
\right), \nonumber \\
\Phi_2 &=& \frac{1}{\sqrt{2}} \left(
\begin{array}{c}
\sqrt{2} (G^+ \sin\beta + H^+ \cos\beta) \\
v\sin\beta + h\cos\alpha + H\sin\alpha + i (G^0 \sin\beta + A \cos\beta)
\end{array} \right),
\end{eqnarray}
where $\tan\beta = \frac{v_2}{v_1}$, and $\alpha$ is the mixing angle between the CP-even states.

The Type-I THDM is characterized by coupling all the standard model (SM) fermions $Q_L$, $u_R$, $d_R$, $L_L$, $e_R$ with the $\Phi_2$ field
\begin{eqnarray}
\mathcal{L}_{\text{Yukawa}}^{\text{SM}} = -Y_{u i j} \overline{Q}_{L i} \tilde{\Phi}_2 u_{R j} - Y_{d i j} \overline{Q}_{L i} \Phi_2 d_{R j} - Y_{l i j} \overline{L}_{L i} \Phi_2 l_{R j} + \text{h.c.},
\end{eqnarray}
where $Y_{u,d,l}$ are the $3 \times 3$ coupling constants. This can be achieved by charging all the right-handed fields with the $-1$, and the left-handed fields with the $+1$ under the $Z_2$ symmetry described above. In the limit that $\tan\beta \rightarrow \infty$ and $\sin(\beta-\alpha) \rightarrow 1$, the couplings between the SM fermions and the exotic Higgs bosons ($H$, $A$, $H^{\pm}$) are highly suppressed by $\sin\alpha$ or $\frac{1}{\tan\beta}$, making them easy to evade various bounds.

Based on the Type-I THDM, if we introduce the sterile neutrinos $N$, and charge them with $+1$ under the $Z_2$ symmetry, we get the $\nu$THDM. In the $\nu$THDM, sterile neutrinos couple with the $L_L$ only through the $\Phi_1$. Since in this paper, we will combine the inverse seesaw mechanisms with the $\nu$THDM, we then introduce three pairs of sterile neutrino fields $N_{Li} = P_L N$, $N_{Ri} = P_R N$ charged with the lepton number 1, where $i=1\text{-}3$, $P_{L,R}=\frac{1 \mp \gamma^5}{2}$, and the Dirac 4-spinors $N_i$ can be written in the form of $\left[ \begin{array}{c} N_{L i}^{\text{w}} \\ i \sigma^2 N_{R_i}^{\text{w} *} \end{array} \right]$. The corresponding Lagrangian is given by
\begin{eqnarray}
\mathcal{L}_{\text{Yukawa}}^{\nu} = -Y_{N i j} \overline{L}_{L i} \tilde{\Phi}_1 N_{R j} - m_{N i j} \overline{N}_{L i} N_{R j} - \mu_{i j} \overline{N_{L i}^c} N_{L j},
\end{eqnarray}
where $Y_N$ is the $3 \times 3$ Yukawa coupling constant matrix, $m_N$ is the $3 \times 3$ Dirac mass matrix between the sterile neutrino pairs, $mu$ is a $3 \times 3$ mass matrix which softly breaks the lepton number, and $N_{L i}^c=-i \gamma^2 \gamma^0 \overline{N_{L i}^c}^T$ is the charge conjugate transformation of the $N_{L i}$ field.

The VEV of the $\Phi_1$ contributes to the Dirac mass terms between the left-handed neutrinos and the sterile neutrinos
\begin{eqnarray}
m_D=\frac{v_1}{\sqrt{2}} Y_N.
\end{eqnarray}
The full $9 \times 9$ mass matrix among the Weyl 2-spinors $\nu_L^{\text{w}}$, $N_L^{\text{w}}$, $N_R^{\text{w}}$ is given by
\begin{eqnarray}
\left[
\begin{array}{ccc}
0 & m_D & 0 \\
m_D^T & 0 & m_N \\
0 & m_N^T & \mu
\end{array}
\right]. \label{Inverse_Seesaw_Mass_Matrix}
\end{eqnarray}
Diagonalizing this matrix gives the light neutrino mass matrix
\begin{eqnarray}
m_{\nu} = m_D m_N^{-1} \mu (m_N^T)^{-1} m_D^T. \label{NeutrinoMass}
\end{eqnarray}
Diagonalizing (\ref{NeutrinoMass}), we need the PMNS matrix
\begin{eqnarray}
& & U = \left[
\begin{array}{ccc}
c_{12} c_{13} & s_{12} c_{13} & s_{13} e^{-i \delta} \\
-s_{12} c_{23} - c_{12} s_{23} s_{13} e^{i \delta} & c_{12} c_{23} - s_{12} s_{23} s_{13} e^{i \delta} & s_{23} c_{13} \\
s_{12} s_{23} - c_{12} c_{23} s_{13} e^{i \delta} & -c_{12} s_{23} - s_{12} c_{23} s_{13} e^{i \delta} & c_{23} c_{13}
\end{array}\right] \times \text{diag}(1, e^{i \frac{\alpha_{21}}{2}}, e^{i \frac{\alpha_{31}}{2}}), \nonumber \\
& & \text{diag}(m_1, m_2, m_3) = U^T m_{\nu} U,
\end{eqnarray}
where $s_{ij}=\sin{\theta_{ij}}$, $c_{ij}=\cos\theta_{ij}$, and $\theta_{ij}$ are the mixing angles, $\delta$ is the CP-phase angle, and $\alpha_{21,31}$ are the two Majorana CP phases. $m_{1,2,3}$ are the masses of the three light neutrinos. Part of the parameters has been measured, and in the rest of this paper, we adopt the following central value \cite{NeutrinoGlobalFit}
\begin{eqnarray}
& & \Delta m_{21}^2 = 7.37 \text{eV}^2, ~~~~~~ |\Delta m^2| = |\Delta m_{32}^2 + \Delta \frac{m_{21}^2}{2}| = 2.50 \text{eV}^2, ~~~~~~ \sin\theta_{12}^2 = 0.297 \nonumber \\
& & \sin^2\theta_{23} = 0.437, ~~~~~~ \sin^2\theta_{13}=0.0214. \label{Neutrino_Parameters}
\end{eqnarray}
We set all the CP phases as zero for simplicity.

To understand the approximate tri-bi-structrue of the $U$ as the $\theta_{13}$ is relatively small compared with other mixing angles, models \cite{InverseFlavon1, InverseFlavon2} have been built by introducing some flavon fields. The Tab.~I in Ref.~\cite{InverseFlavon1} listed seven cases of different $m_D$, $m_N$, $\mu$ combinations in such kind of models. In this paper, we only discuss the previous three cases. They are listed in Tab.~\ref{Combination_Tribi}. Unlike Ref.~\cite{InverseFlavon1}, here $M_0$ should be compatible with a non-zero $\theta_{13}$, just as the example revealed in Ref.~\cite{InverseFlavon2}. 
\begin{table}
\begin{tabular}{|c|c|c|c|}
\hline
cases & 1) & 2) & 3) \\
\hline
$m_D$ & $M_0$ & $\propto I$ & $\propto I$ \\
\hline
$m_N$ & $\propto I$ & $M_0$ & $\propto I$ \\
\hline
$\mu$ & $\propto I$ & $\propto I$ & $M_0$ \\
\hline
\end{tabular}
\caption{Possible $m_D$, $m_N$, $\mu$ combinations. Here $M_0$ means a matrix which is not proportional to the identical matrix $I$. } \label{Combination_Tribi}
\end{table}

Define
\begin{eqnarray}
m_{\nu}^{\frac{1}{2}} = U \cdot \text{diag}(\sqrt{m_1}, \sqrt{m_2}, \sqrt{m_3}) \label{mnu_Half}
\end{eqnarray}
so that $m_{\nu}^{\frac{1}{2}} (m_{\nu}^{\frac{1}{2}})^T = m_{\nu}$. Therefore, during the numerical calculation processes, we set
\begin{eqnarray}
m_D \propto m_{\nu}^{\frac{1}{2}}, ~~~~ m_N \propto I, ~~~~ \mu \propto I \label{Case_I}
\end{eqnarray}
in the case 1),
\begin{eqnarray}
m_D \propto I, ~~~~ m_N \propto (m_{\nu}^{\frac{1}{2}})^T, ~~~~ \mu \propto I \label{Case_II}
\end{eqnarray}
in the case 2), and
\begin{eqnarray}
m_D \propto I, ~~~~ m_N \propto I, ~~~~ \mu \propto m_{\nu} \label{Case_III}
\end{eqnarray}
in the case 3). Note that the definition in (\ref{mnu_Half}) of the $m_{\nu}^{\frac{1}{2}}$ is not the only one that can reach $m_{\nu}^{\frac{1}{2}} (m_{\nu}^{\frac{1}{2}})^T = m_{\nu}$. However, all the other definitions can be equavalent with the (\ref{mnu_Half}) by redefining the $N_{L,R}$ fields, so it is enough to adopt (\ref{Case_I}-\ref{Case_III}) in all the three cases.

\section{Calculations of the Observables}

The Z-boson mass $m_Z$, the Fermi constant $G_F$ and the fine structure constant $\alpha$ are the three parameters with the smallest experimental errors. Together with the strong coupling constant $\alpha_s$, the SM-Higgs boson mass $m_h$, and the fermion masses and mixings, these parameters can be used as the input parameter set to evaluate other observables. Ref.~\cite{PDG} states that their fits of the ``SM-values'' are not the pratical consequences for the precisely known $\alpha$, $G_F$ and $m_Z$. However, In principle we can always calculate the ``SM-predicted'' values of the observables from the parameters listed above, and compare them with the measured ones on various (proposed future) experiments.

In this paper, we mainly discuss about Z-resonance observables They are $R_l = \frac{\Gamma_{Z \rightarrow \text{hadrons}}}{\Gamma_{Z \rightarrow l^{+} l^{-}}}$, $A_l = \frac{2 \overline{g}_V^l \overline{g}_A^l}{\overline{g}_V^{l 2}+\overline{g}_A^{l 2}}$. The muon anomolous $g-2$, the lepton's FCNC decay $\tau \rightarrow e/\mu+\gamma$, $\mu \rightarrow e + \gamma$ are also calculated. All the SM input parameters can be measured independently from these observables. For example, the Fermi constant $G_F$ can be extracted from the precisely measured muon mass and its lifetime \cite{MuonLifetime}, and the current value of the fine structure constant $\alpha$ originate from low-energy experiments, and the  $\widehat{\alpha}(m_Z)$ defined in the modified minimal subtraction ($\overline{\text{MS}}$) is then calculated considering the vacuum polarization effects of the leptons and hadrons (In \cite{PDG}, there is a review, and for the references therein) . Another example is the $\alpha_s$, which can be extracted from the $R_l$, though, there are various other measures to acquire its value which can reach at least similar precisions.

In some cases, the new physics sectors might shift the values of the SM input parameters, altering the ``SM-predicted'' values of some observables. In this paper, we should note that the decay width $\Gamma_{\mu \rightarrow e \nu \overline{\nu}}$ can be affected by the $H^{\pm}$ mediator, shifting the measured fermi constant $G_F$ from its ``real value''. We consider this effect in our following discussions, however, we do not care about the breaking of lepton universality of the ``flavorful'' gauge couplings $g_{e,\mu,\tau}$ (For an example, see Ref.~\cite{vTHDM_Dirac2}. See Ref.~\cite{Tau_Break_Universality} for the experimental results) at the moment in this paper.

In order to calculate the shift of the decay width of the muon, we need to diagonalize the $m_N$ matrix beforehand. Suppose $m_N$ has been diagonalized, and $m_N^i$'s are the eigenvalues of this matrix, then the shift to the muon's decay width is given by \cite{vTHDM_Dirac1, ModifyGF}
\begin{eqnarray}
\Gamma_{\mu} = \Gamma_{\mu, \text{SM}} \left[ 1+ \left(\frac{v}{\sqrt{2} m_{H^{\pm}}^4}\right)^4 \frac{\displaystyle{\sum_{i=1\text{-}3, l=e,\mu,\tau}} U^{\nu N}_{l i} Y_{N e i} \displaystyle{\sum_{j=1\text{-}3, l^{\prime}=e,\mu,\tau}^3} U^{\nu N}_{l^{\prime} j}Y_{N \mu j}}{4} \right],
\end{eqnarray}
where $U^{\nu N}$ is the mixing between the light neutrinos and the sterile neutrinos when diagonlizing (\ref{Inverse_Seesaw_Mass_Matrix}). Then the shift of the $G_F$ can be estimated as
\begin{eqnarray}
G_F &\rightarrow& G_F + \delta G_F, \nonumber \\
\delta G_F &\approx& G_F  \left(\frac{v}{\sqrt{2} m_{H^{\pm}}^4}\right)^4 \frac{\displaystyle{\sum_{i=1\text{-}3, l=e,\mu,\tau}} U^{\nu N}_{l i} Y_{N e i} \displaystyle{\sum_{j=1\text{-}3, l^{\prime}=e,\mu,\tau}^3} U^{\nu N}_{l^{\prime} j}Y_{N \mu j}}{8}.
\end{eqnarray}
The values of the $U^{\nu N}$'s are calculated to be
\begin{eqnarray}
U^{\nu N}_{l, i}= - \frac{Y_{N l i} v \cos\beta}{m_{N i}}.
\end{eqnarray}
Notice that some of the tree-level definitions of the electroweak observables are functions depending only on the weak mixing angle $\theta_W$. Therefore, we need to calculate the shifting of the $\theta_W$,
\begin{eqnarray}
& & \frac{8 G_F M_Z}{\sqrt{2} e^2} = \frac{1}{\sin^2 \theta_W \cos \theta_W}, \nonumber \\
&\rightarrow & \delta \theta_W = \frac{8 \delta G_F M_Z}{\sqrt{2} e^2} \left(\frac{-2}{\sin^3 \theta_W} + \frac{1}{\sin \theta_W \cos^2 \theta_W} \right). \label{thetaW_shifting}
\end{eqnarray}

Now we are ready to calculate the
\begin{eqnarray}
\delta R_l &=& R_l^{\text{exp.}} - R_l^{\text{SM Pre.}}, \nonumber \\
\delta A_l &=& A_l^{\text{exp.}} - A_l^{\text{SM Pre.}}, \nonumber \\
\delta N_{\nu} &=& N_{\nu}^{\text{exp.}} - 3,
\end{eqnarray}
where
\begin{eqnarray}
R_{l_i} &=& \frac{\Gamma_{Z \rightarrow \text{had}}}{\Gamma_{Z \rightarrow l_i^+ l_i^-}} \nonumber \\
A_{l_i} &=& \frac{2 \overline{g}_V^{l_i} \overline{g}_A^{l_i}}{\overline{g}_V^{l_i 2}+ \overline{g}_A^{l_i 2}},
\end{eqnarray}
and the superscript ``exp.'', ``SM Pre.'' indicate the experimentally measured values and the ``SM-predicted'' values considering the shifting of the Fermi constant $G_F$.
The definitions of the $N_\nu$ are a little bit complicated, and will be discussed later. All of the $\delta X$'s involve the corrections to the effective coupling constants $\overline{g}^{ffZ}_{A,V, L, R}$'s defined by 
\begin{eqnarray}
\mathcal{L}_{ffZ} & = & \frac{-e}{2 \sin \theta_W \cos \theta_W} Z_{\mu} \overline{f} \gamma^{\mu} [ \overline{g}^f_L \frac{1-\gamma^5}{2} + \overline{g}^f_R \frac{1+\gamma^5}{2} ] f \nonumber \\
& = & \frac{-e}{2 \sin \theta_W \cos \theta_W} Z_{\mu} \overline{f} \gamma^{\mu} ( \overline{g}^f_V - \overline{g}^f_A \gamma^5 ) f,
\end{eqnarray}
where
\begin{eqnarray}
\overline{g}^f_V = \overline{g}^f_L+\overline{g}^f_R,~~~~\overline{g}^f_A=\overline{g}^f_L-\overline{g}^f_R,
\end{eqnarray}
and
\begin{eqnarray}
\overline{g}^{f}_{L,R,V,A} = g^{f}_{L,R,V,A}+\delta g^{f}_{L,R,V,A},
\end{eqnarray}
where $g^{f}_{L,R,V,A}$ are the SM values, and the $\delta g^{f}_{L,R,V,A}$ are the new physics corrections.

\begin{figure}
  \includegraphics[width=1.5in]{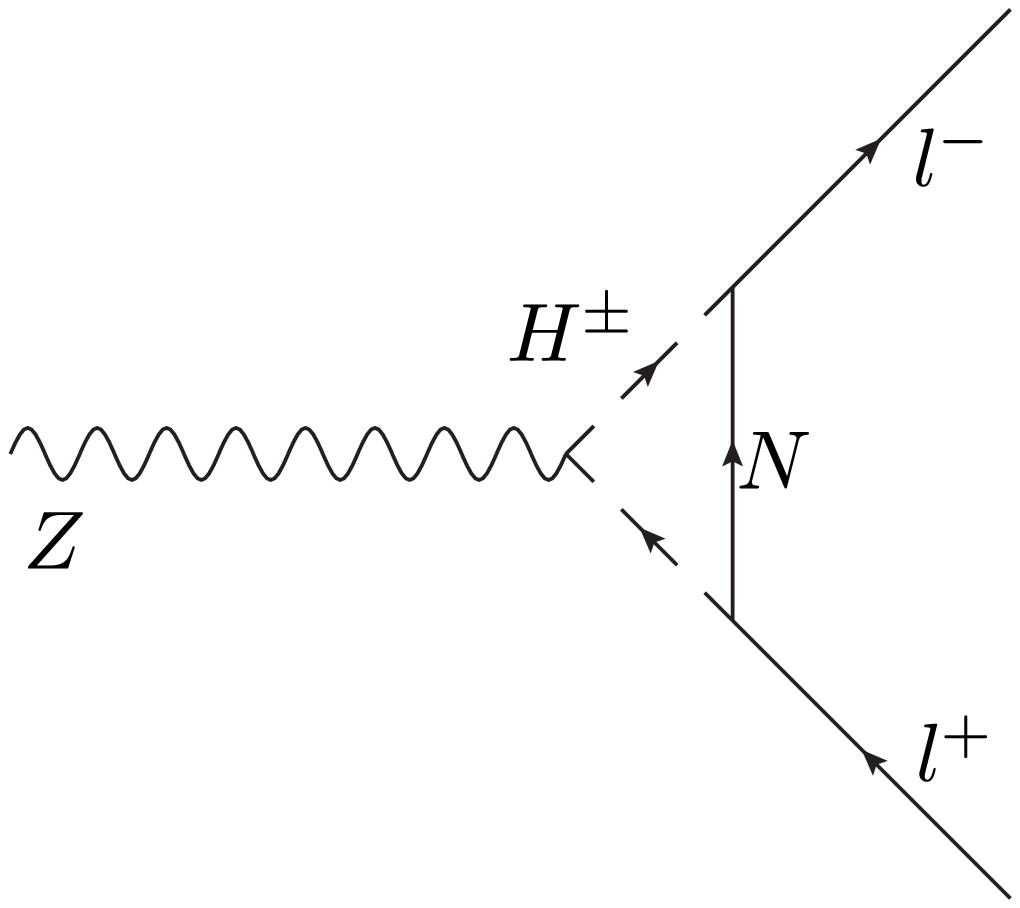}
  \caption{(a) Diagrams to the $Z$-$l^+$-$l^0$ vertices. \label{Zll_a}}
  \includegraphics[width=3in]{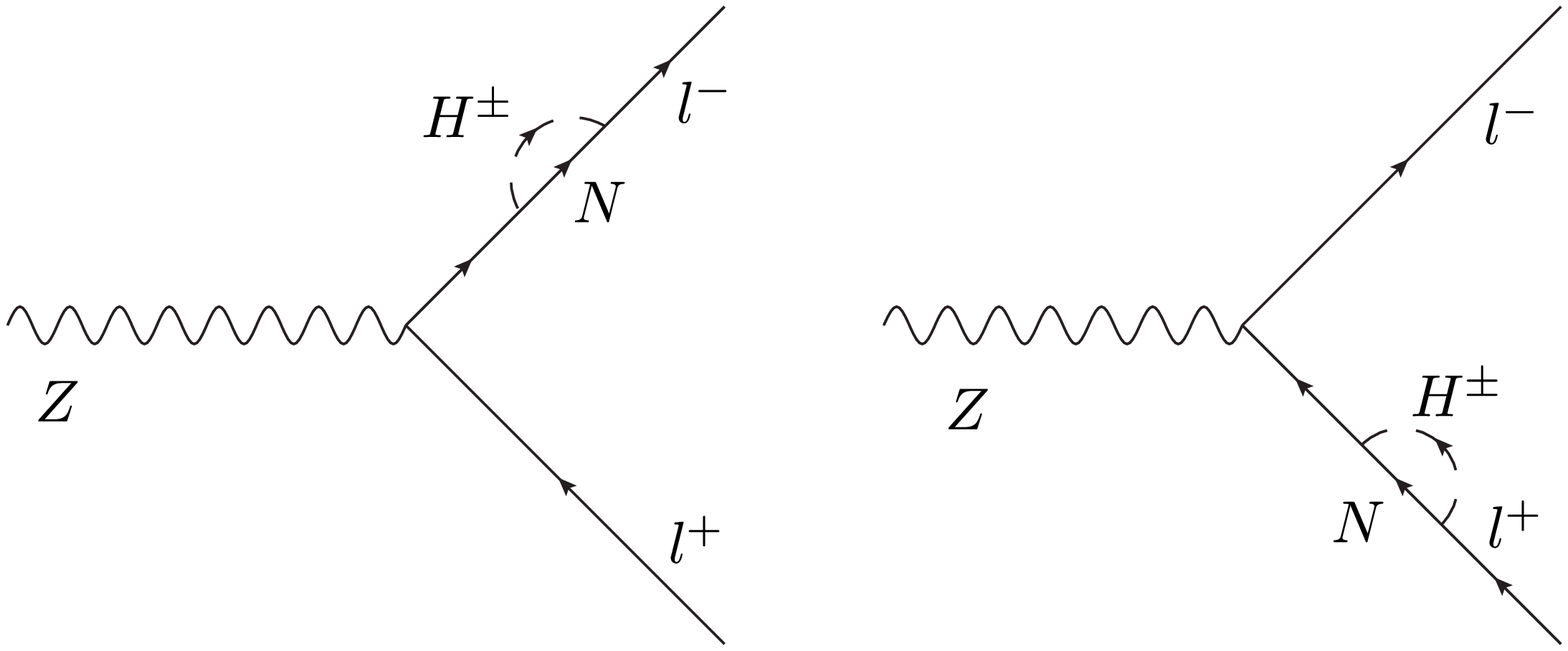}
  \caption{(c) Diagrams to $l^{\pm}$ propagators. \label{Zll_c}}
\end{figure}

To calculate the $Z$-$l^{+}$-$l^{-}$ loop corrections where $l=e, \mu, \tau$, we need to calculate the Feynmann diagrams in Fig.~\ref{Zll_a}, \ref{Zll_c}. The Ref.~\cite{Zbb_Ancester} had calculated the loop corrections to the $Z$-$b$-$b$ vertices, and it is easy to modify the formulas there to evaluate the $Z$ vertices in this paper. suppose $m_N$ have been diagonalized, we have
\begin{eqnarray}
\delta g_L^{l_1 l_2 (a)} &=& \frac{1}{8 \pi^2} Y_{N l_1 j} Y^{*}_{N l_2 j} g_L^{Z H^+ H^-} C_{00} (0, 0, m_Z^2, m_{H^{\pm}}^2, m_{N i}^2, m_{H^{\pm}}^2), \nonumber \\
\delta g_L^{l_1 l_2 (c)} &=& \frac{1}{16 \pi^2} y_{N l_1 j} Y^{*}_{N l_2 j} g_L^{Z l_1^+ l^-} B_1 (0, |m_{N i}|^2, m_{H^{\pm}}^2), \nonumber \\
\delta g_L^{l_1 l_2} &=& \delta g_L^{l_1 l_2 (a)}+\delta g_L^{l_1 l_2 (c)}, \nonumber \\
\delta g_R^{l_1 l_2} &=& 0, \label{Charged_Lepton_Corrections}
\end{eqnarray}
for lepton $l_1$, and $l_2$. $C_{ij}$, $B_i$ are the Passarino-Veltman integrals with the conventions of the parameters similar to the LoopTools manual \cite{LoopTools}. We also ignore all the leptonic masses during the calculations. Notice that if $l_1 \neq l_2$, the (\ref{Charged_Lepton_Corrections}) can result in a FCNC $Z \rightarrow l_1 l_2$ decay. In this paper, we are not going to talk about them since they are exceeding the abilities of many collider experiments.

\begin{figure}
  \includegraphics[width=3in]{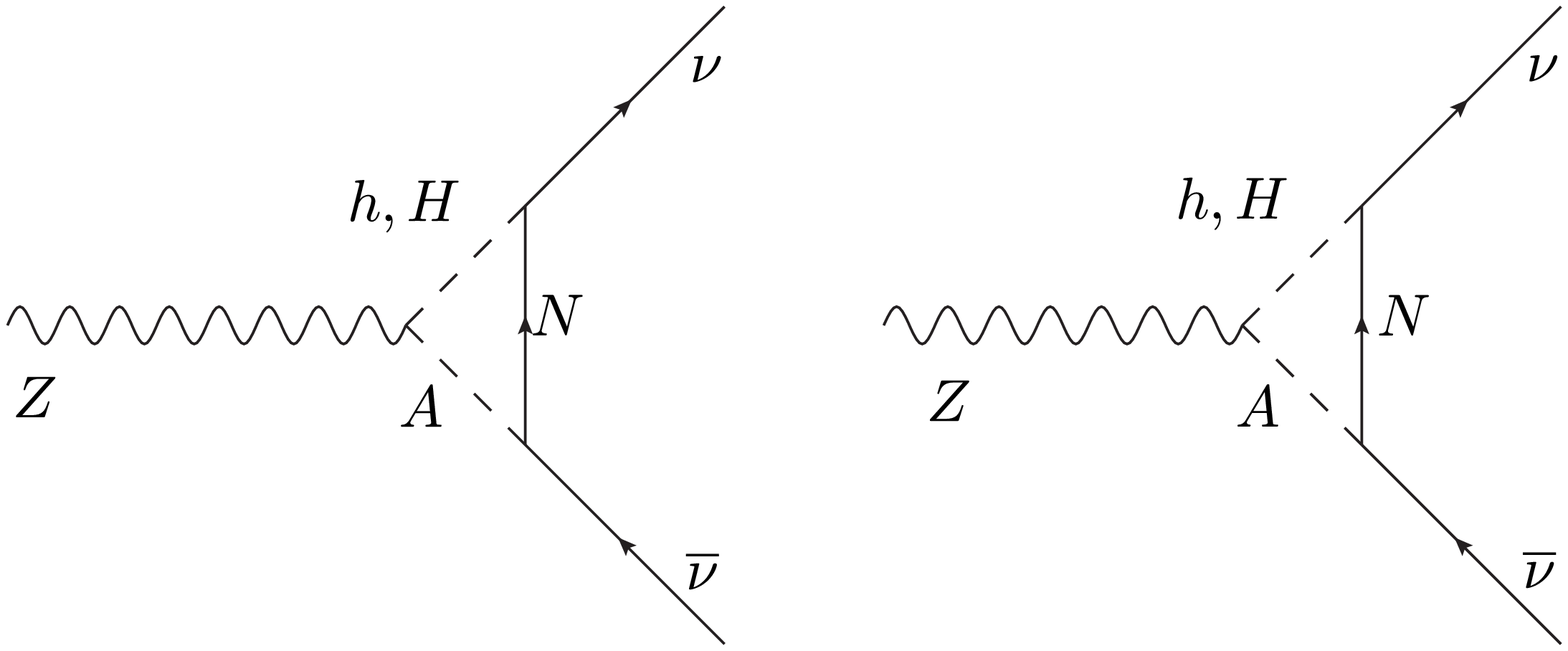}
  \caption{(a) Diagrams to the $Z$-$\nu$-$\overline{\nu}$ vertices. \label{Zvv_a}}
  \includegraphics[width=3in]{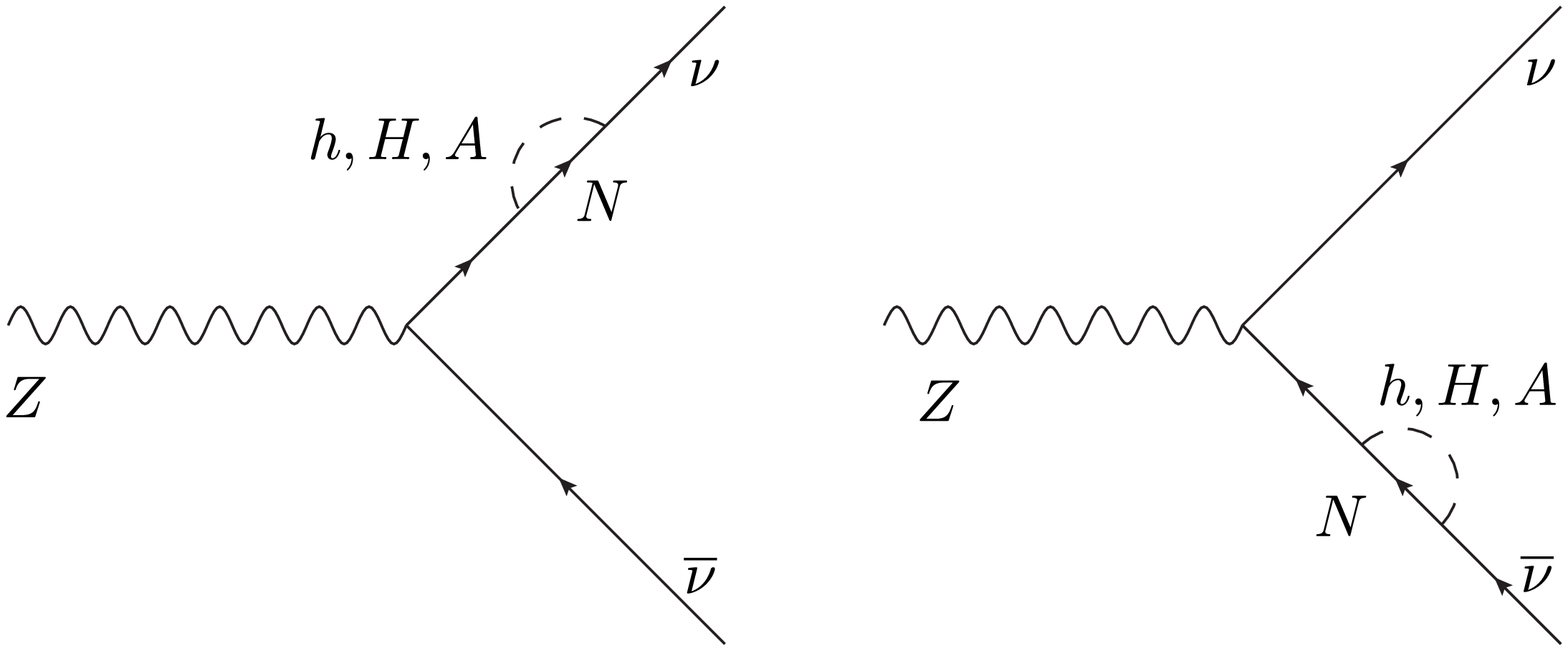}
  \caption{(c) Diagrams to $\nu$ propagators. \label{Zvv_c}}
\end{figure}

The $Z \rightarrow \nu \nu$ vertices also receive loop corrections. By calculating the Feynmann diagrams in Fig.~\ref{Zvv_a}, \ref{Zvv_c}, we have
\begin{eqnarray}
\delta g_L^{\nu_{l 1} \nu_{l 2} (a)} &=& -\frac{1}{8 \pi^2} g^{ZhA} \displaystyle{\sum_i} (y^{\nu N h}_{l_1 i} y^{\nu N A *}_{l_2, i} + y^{\nu N h *}_{l_1 i} y^{\nu N A}_{l_2, i}) C_{00} (0, 0, m_Z^2, m_h^2, |m_{N i}|^2, m_A^2) \nonumber \\
&-& \frac{1}{8 \pi^2} g^{ZHA} \displaystyle{\sum_i} (y^{\nu N H}_{l_1 i} y^{\nu N A *}_{l_2, i} + y^{\nu N H *}_{l_1 i} y^{\nu N A}_{l_2, i}) C_{00} (0, 0, m_Z^2, m_H^2, |m_{N i}|^2, m_A^2), \nonumber \\
\delta g_L^{\nu_{l 1} \nu_{l 2} (c)} &=& \frac{1}{32 \pi^2} g_L^{Z \nu \nu} \displaystyle{\sum_i} ( y^{\nu N h}_{l_1 i} y^{\nu N h *}_{l_2 i} + y^{\nu N h *}_{l_1 i} y^{\nu N h}_{l_2 i} ) B_1 (0, |m_{N i}|^2, m_h^2) \nonumber \\
&+& \frac{1}{32 \pi^2} g_L^{Z \nu \nu} \displaystyle{\sum_i} ( y^{\nu N H}_{l_1 i} y^{\nu N H *}_{l_2 i} + y^{\nu N H *}_{l_1 i} y^{\nu N H}_{l_2 i} ) B_1 (0, |m_{N i}|^2, m_H^2) \nonumber \\
&+& \frac{1}{32 \pi^2} g_L^{Z \nu \nu} \displaystyle{\sum_i} ( y^{\nu N A}_{l_1 i} y^{\nu N A *}_{l_2 i} + y^{\nu N A *}_{l_1 i} y^{\nu N A}_{l_2 i} ) B_1 (0, |m_{N i}|^2, m_A^2), \nonumber \\
\delta g_{L \text{loop}}^{\nu_{l 1} \nu_{l 2}}&=&\delta g_L^{\nu_{l 1} \nu_{l 2} (a)} + \delta g_L^{\nu_{l 1} \nu_{l 2} (c)} \label{delta_gv_loop}
\end{eqnarray}
where $y^{\nu N (h, H, A)}_{l_i j}$ are the $\nu$-$N$-neutral Higgs coupling constants after everything is rotated to their mass eigenstates. 

In Fig.~\ref{Zll_a}-\ref{Zvv_c}. We name the diagram sets ``(a)'' and ``(c)'' in order to compare our diagrams and results with the Ref.~\cite{Zbb_Ancester}, and we should note that the ``(b)'', ``(d)'', etc., are absent because $N_i$ are SM neutral particles. In the Fig.~\ref{Zll_a}-\ref{Zll_c}, sterile neutrino propagators are with arrows since they are pseudo-Dirac particles, and the corrections involving $\mu$ are ommited.

Despite the loop corrections to the $Z \rightarrow \overline{\nu} \nu$ vertices, tree-level shiftings due to the mixings between the light neutrinos and the sterile neutrinos should also be considered. Up to the lowest order,
\begin{eqnarray}
\delta g_{L \text{tree}}^{\nu_{l 1} \nu_{l 2}} = - \delta_{\nu_{l 1} \nu_{l 2}} g_L^{Z \nu \nu} \displaystyle{\sum_{i}} \frac{m_{D l_1 i}^2}{2 m_{N i}^2}. \label{delta_gv_tree}
\end{eqnarray}
In our numerical evaluations, both (\ref{delta_gv_loop}) and (\ref{delta_gv_tree}) are considered.

The definitions of the $R_l$, $A_l$, and the $N_{\nu}$ are some ratios among expressions of $\overline{g}_{L,R}^{f f}$, or equivalently, $\overline{g}_{V,A}^{f f}$. Here $f f$ include all the lepton and quark pairs. In the model discussed in this paper, the new-physics corrections to the $Z$-quarks couplings from the SM values can be ignored. We also ommit the SM-radiative corrections during our evaluations since we only pay attentions to the new physics effects. Then, $\delta R_l$, $\delta A_l$ are given by
\begin{eqnarray}
\delta R_{l_i} &=& -\frac{4 (-19 \sin 2 \theta_W + 14 \sin 4 \theta_W + 5 \sin 6 \theta_W)}{3 (2 - 2 \cos 2 \theta_W + \cos 4 \theta_W )^2} \delta \theta_W \nonumber \\
&+& \frac{2 (-38 + 85 \cos 2 \theta_W - 13 \cos 4 \theta_W + 11 \cos 6 \theta_W )}{3 ( 2 - 2 \cos 2 \theta_W + \cos 4 \theta_W)^2} \delta g_V^{l_i l_i} \nonumber \\
&+& \frac{2 ( 36 - 2 \cos 2 \theta_W + 11 \cos 4 \theta_W)}{3 (2 - 2 \cos 2 \theta_W + \cos 4 \theta_W)^2} \delta g_A^{l_i l_i}, \label{delta_R_l}  \\
\delta A_{l_i} &=& \frac{8 \sin^2 \theta_W \sin 4 \theta_W}{(2 - 2 \cos 2 \theta_W + \cos 4 \theta_W)^2} \delta \theta_W \nonumber \\
&-& \frac{8 \cos 2 \theta_W \sin^2 \theta_W}{(2-2 \cos 2 \theta_W + \cos 4 \theta_W)^2} \delta g_V^{l_i l_i} \nonumber \\
&+& \frac{8 (1-\cos 2 \theta_W + \cos 4 \theta_W) \sin^2 \theta_W}{(2 - 2 \cos 2 \theta_W + \cos 4 \theta_W)^2} \delta g_A^{l_i l_i}, \label{delta_A_l}
\end{eqnarray}
where the first terms in both (\ref{delta_R_l}) and (\ref{delta_A_l}) originate from the shifting of the $G_F$, while the rest of the terms indicate the radiative corrections from the charged Higgs loops.

As for the $\delta N_{\nu}$, things are a little bit subtle. The definition given by the Ref.~\cite{PDG} is
\begin{eqnarray}
N_{\nu}^l = \frac{\Gamma^Z_{\text{inv}}}{\Gamma^Z_l} \left( \frac{\Gamma^Z_l}{\Gamma^Z_{\nu}} \right)_{\text{SM}},
\end{eqnarray}
where the $\left(\frac{\Gamma^Z_{\nu}}{\Gamma^Z_{l}}\right)_{\text{SM}}$ is used instead of $(\Gamma_{\nu})_{\text{SM}}$ in order to reduce the model dependence. However, in our model, both $\Gamma_{l}$ and $\Gamma_{\nu}$ receive corrections. We also define and will calculate the
\begin{eqnarray}
N_{\nu}^{\text{h}} = \frac{\Gamma^Z_{\text{inv}}}{\Gamma^Z_{\text{h}}} \left( \frac{\Gamma^Z_{\text{h}}}{\Gamma^Z_{\nu}} \right)_{\text{SM}},
\end{eqnarray}
where $\Gamma^Z_{\text{h}}$ is the partial width that $Z$ boson decays to hadrons, for comparison, since $Z$-hadrons couplings do not receive significant new physics corrections in this model. They are given by
\begin{eqnarray}
\delta N_{\nu}^{l} & = & \frac{12 (\sin 2 \theta_W - \sin 4 \theta_W)}{2-2 \cos 2 \theta_W + \cos 4 \theta_W} \delta \theta_W + 2 \displaystyle{\sum_{i}} (\delta g_V^{\nu_{l i} \nu_{l i}} + \delta g_A^{\nu_{l i} \nu_{l i}}) \nonumber \\
& + & \frac{2-8 \sin^2 \theta_W}{2-2 \cos 2 \theta_W + \cos 4 \theta_W}  \displaystyle{\sum_i} \delta g_V^{l_i l_i} + \frac{2}{2-2 \cos 2 \theta_W + \cos 4 \theta_W} \displaystyle{\sum_i} \delta g_A^{l_i l_i}, \label{delta_Nv_l} \\
\delta N_{\nu}^{\text{h}} & = & \frac{12 (\sin 2 \theta_W - 11 \sin 4 \theta_W)}{36 - 2 \cos 2 \theta_W + 11 \cos 4 \theta_W} \delta \theta_W + 2 \displaystyle{\sum_{i}} (\delta g_V^{\nu_i \nu_i} + \delta g_A^{\nu_i \nu_i}), \label{delta_Nv_h}
\end{eqnarray}
where $\delta g_{L,R,V,A}^{l_i l_j} = \delta g_{L,R,V,A\text{tree}}^{l_i l_j} + \delta g_{L,R,V,A\text{loop}}^{l_i l_j}$, and again the first terms in both (\ref{delta_Nv_l}) and (\ref{delta_Nv_h}) originate from the shifting of the $G_F$ while the other terms come from the corrections to the effective $Z$-$f$-$\overline{f}$ corrections, containing both the tree-level and the loop-level ones.

We should note that strictly speaking, the ``$\theta_W$'' in the (\ref{delta_R_l}-\ref{delta_Nv_h}) should be replaced by ``$\arcsin(s_l)$'', which is the angle evaluated from the SM-effective $Z$-$l$-$l$ vertices. However, in this paper, we only concern the deviations from the SM predictions, which is insensitive to the definitions of the weak mixing angle, so we do not distinguish them.

\begin{figure}
\includegraphics[width=3in]{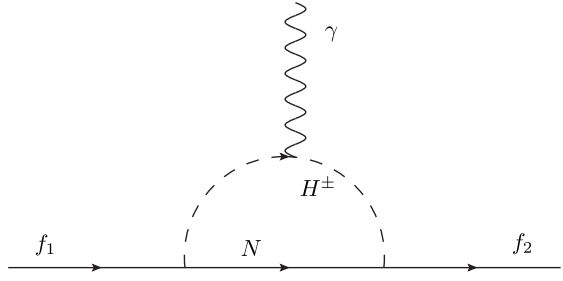}
\caption{The diagram for $l_1 \rightarrow l_2 \gamma$. This diagram can also be used to calculate the muon anomalous $g-2$.} \label{l1l2gamma}
\end{figure}

The lepton's FCNC decay $\mu \rightarrow e \gamma$, $\tau \rightarrow \mu \gamma$, $\tau \rightarrow e \gamma$ processes together with the muon anomalous $g-2$ provide other windows towards the new physics models. All of them involve a one-loop diagram with a charged Higgs boson running inside. The diagram is shown in Fig.~\ref{l1l2gamma}. We follow the steps in Ref.~\cite{f1f2gamma} to calculate the amplitute, which is parametrized by $i e \epsilon_{\mu}^*(q) M^{\mu}$, where $e=\sqrt{4 \pi \alpha}$ is the coupling constant of the quantum electromagnetic dynamics, $\epsilon_{\mu}^q$ is the polarization vector. The definition of the $M^{\mu}$ is given by
\begin{eqnarray}
M^{\mu} = \overline{u}_2 [ i \sigma^{\mu \nu} q_\nu (\sigma_{L l_1 l_2} P_L + \sigma_{R l_1 l_2} P_R) u_1, \label{l1l2gamma_Amplitute}
\end{eqnarray}
where $P_{L, R} = \frac{1 \mp \gamma^5}{2}$, and $\sigma^{\mu \nu} = \frac{i [ \gamma^{\mu}, \gamma^{\nu}]}{2}$. If $l_1 \neq l_2$, the partial width for $f_1 \rightarrow f_2 \gamma$ is given by
\begin{eqnarray}
\Gamma_{l_1 \rightarrow l_2 \gamma} = \frac{(m_{l_1}^2 - m_{l_2}^2)^3 (|\sigma_{L l_1 l_2}|^2 + |\sigma_{R l_1 l_2}|^2)}{16 \pi m_{l_1}^3}. \label{FCNC_Decay}
\end{eqnarray}
If $l_1 = l_2$, (\ref{l1l2gamma_Amplitute}) also contributes to the anomaly magnetic momenta
\begin{eqnarray}
\delta a_{l_1} = \frac{\sigma_{L l_1 l_1} + \sigma_{R l_1 l_1}}{\frac{e}{2 m_{l_1}}}. \label{Muon_Anomaly_Magnetic_Momenta}
\end{eqnarray}

Define
\begin{eqnarray}
t_i &=& \frac{m_{N i}^2}{m_{H^{\pm}}^2}, \nonumber \\
\overline{c}_{1, i} &=& \overline{c}_{2, i} = \frac{1}{16 \pi^2 m_{H^\pm}^2} \left[ \frac{3 t_i-1}{4 (t_i-1)^2} - \frac{t_i^2 \ln t_i}{2 (t_i-1)^3} \right], \nonumber \\
\overline{d}_{1, i} &=& \overline{d}_{2, i} = 2 \overline{f}_i = \frac{1}{16 \pi^2 m_{H^{\pm}}^2} \left[ \frac{11 t_i^2 - 7 t_i + 2}{18 (t_i-1)^3} - \frac{t_i^3 \ln t_i}{3 (t_i-1)^4} \right], \nonumber \\
\lambda_{l_1 l_2 i} &=& y_{N l_2 i}^* y_{N l_1 i}, \text{~(No Einstein summation rules for the index $i$)}, \nonumber \\
\overline{k}_{1, l_1 l_2 i} &=& m_{l_1} ( - \overline{c}_1 + \overline{d}_1 + \overline{f} ), \nonumber \\
\overline{k}_{2, l_1 l_2 i} &=& m_{l_2} ( - \overline{c}_2 + \overline{d}_2 + \overline{f} ).
\end{eqnarray}
Then the $\sigma_{L,R l_1 l_2}$ are given by
\begin{eqnarray}
\sigma_{L, l_1, l_2} &=& Q_B \lambda \overline{k}_2, \nonumber \\
\sigma_{R, l_1, l_2} &=& Q_B \lambda \overline{k}_1. \label{sigma_LR_Result}
\end{eqnarray}
By taking (\ref{sigma_LR_Result}) to (\ref{FCNC_Decay}-\ref{Muon_Anomaly_Magnetic_Momenta}), we can then calculate the partial widths of the FCNC decay of the $\mu \rightarrow e \gamma$, $\tau \rightarrow \mu \gamma$, $\tau \rightarrow e \gamma$ processes together with the muon anomalous $g-2$.

\section{Numerical Calculations and Results} \label{Numerical_Results}

In this section, we are going to show the results of the $\delta R_{l_i}$, $\delta A_{l_i}$, $\delta N_{\nu}$ together with the bounds from $\mu \rightarrow e \gamma$, $\tau \rightarrow \mu \gamma$, $\tau \rightarrow e \gamma$ in each case listed in Tab.~\ref{Combination_Tribi}. The muon's anomalous $g-2$ is also considered.

Since we mainly concern the Z-resonance observables involving the leptons, the interactions among the Higgs sectors are less important. Under the $\tan \beta \rightarrow \infty$ limit and the alignment limit $\sin ( \beta - \alpha) \rightarrow 1$, only the mass sepctrum of the Higgs bosons and the sterile neutrinos, together with their Yukawa coupling constants play the key roles in resolving the observables. The spectrum of the sterile neutrinos and their Yukawa couplings are affected by the left-handed neutrino mass spectrum and their mixing patterns. After adopting the data in (\ref{Neutrino_Parameters}) and ignoring all the CP phases, we still need the lightest neutrino mass $m_{\nu 0}$ to determine the complete neutrino mass spectrum. Both the normal ordering $m_1<m_2<m_3$ and the inverse ordering $m_3 < m_1 < m_2$ are calculated, however only the results for the normal ordering are presented since there is no significant defferent between these two orderings.

Despite the light neutrino mass and mixing parameters, $m_N$, $m_D$ can be characterised by the lightest sterile neutrino's mass $m_{N 0}$, and the largest SM-effective $y_{\text{SM}}^{\text{max}}$. The $y_{\text{SM}}^{\text{max}}$ is defined by the value of the element with the smallest absolute value in the SM-effective coupling matrix $Y_N \cos \beta$. Besides, $m_{H^\pm}$ determines the $R_{l_i}$ and $A_{l_i}$, while $m_{H,A}$ also affect the $N_{\delta \nu}$. In this paper, we fix $m_h = 125 \text{ GeV}$.

As for the $l_1 \rightarrow l_2 \gamma$ bounds, we adopt the data from Ref.~\cite{mu2egamma, ta2emugamma, PDG},
\begin{eqnarray}
\text{Br}_{\mu \rightarrow e \gamma} &<& 5.7 \times 10^{-13}, \nonumber \\
\text{Br}_{\tau \rightarrow \mu \gamma} &<& 4.4 \times 10^{-8}, \nonumber \\
\text{Br}_{\tau \rightarrow e \gamma} &<& 3.3 \times 10^{-8}.
\end{eqnarray}
The Planck collaboration also gives constraints on the summation of the light neutrino mass \cite{PlanckNeutrino}
\begin{eqnarray}
\displaystyle{\sum_i} m_{\nu_{i}} &<& 0.23 \text{ eV}.
\end{eqnarray}
The deviation of the muon's anamous magnetic momenta between the experimental and the theoretical evaluation results is $\delta a_{\mu} = 288(63)(49) \times 10^{-11}$ \cite{Muongm2_1, Muongm2_2, Muongm2_3, PDG}. Here we adopt the $3-\sigma$ range of
\begin{eqnarray}
48.56 \times 10^{-11} < \delta a_{\mu} < 527.44 \times 10^{-11}.
\end{eqnarray}
Since in many cases, the differences between the $\delta N_{\nu}^{l}$ and the $\delta N_{\nu}^{\text{h}}$ are not very significant, we refer to the $\delta N_{\nu}^l$ when we refer to $\delta N_{\nu}$.

\begin{figure}
\includegraphics[width=0.45\textwidth]{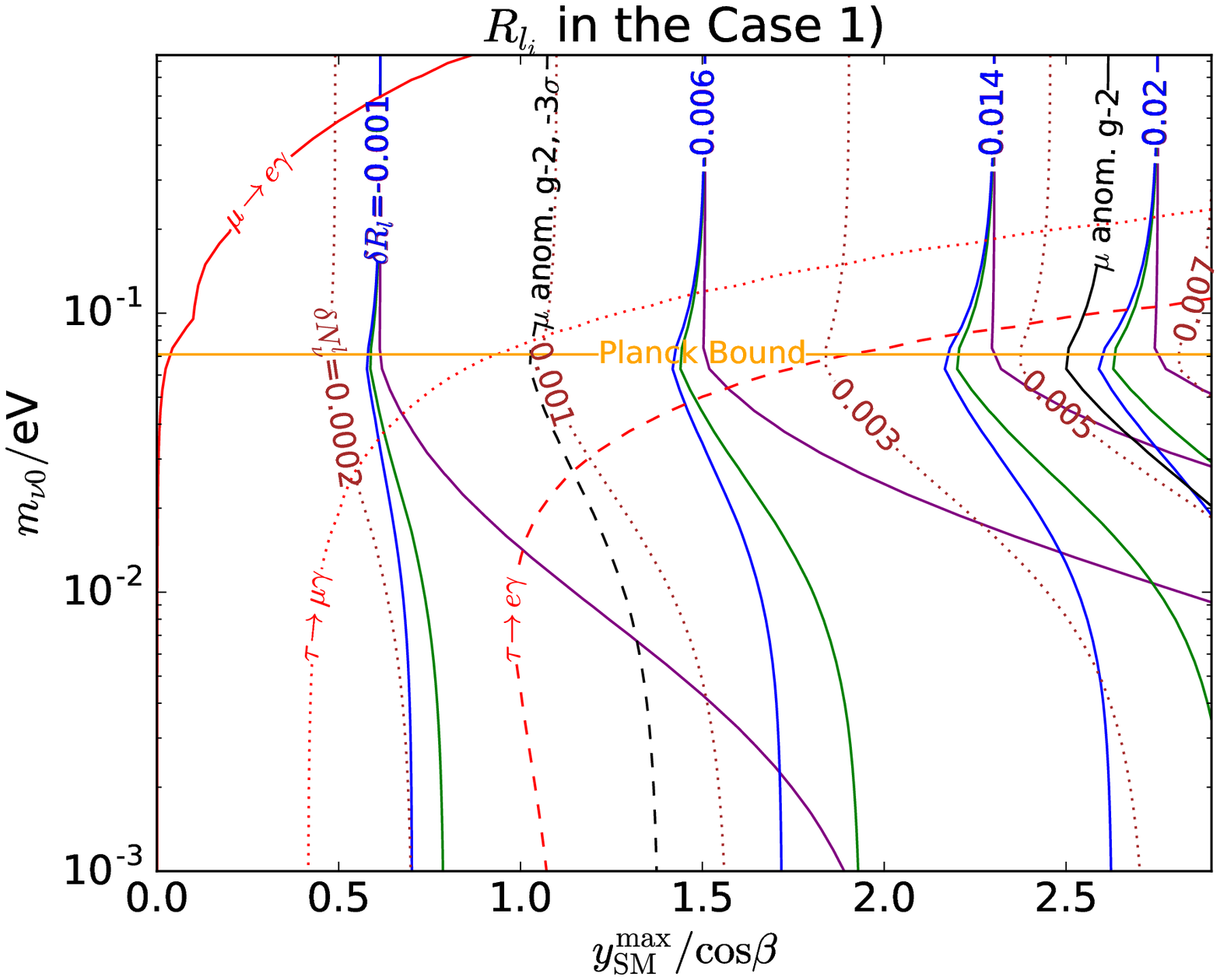}
\includegraphics[width=0.45\textwidth]{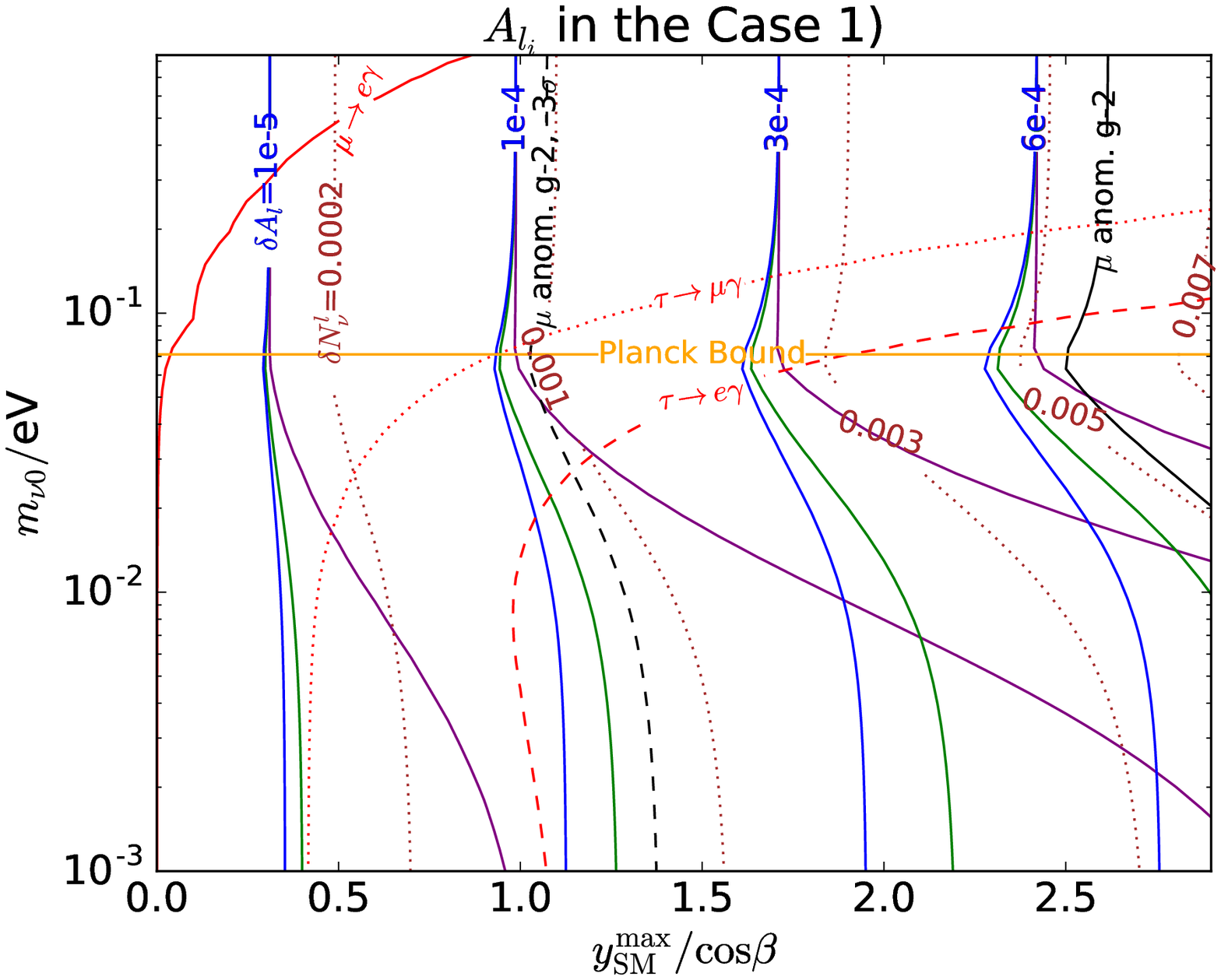}
\caption{The $R_{l_i}$ (left panel) and $A_{l_i}$ (right panel) together with the $l_1 \rightarrow l_2 \gamma$ bounds and the 3-$\sigma$ $g-2$ range. Here $m_H = m_{H^{\pm}} = m_{A} = 200 \text{ GeV}$. $\tan \beta=1000$, $\sin (\beta-\alpha) = 0.9999$, and $m_{N_0}=20 \text{ GeV}$. The purple, green, blue lines indicate the $\delta R/\delta A_{e,\mu,\tau}$ respectively.} \label{Normal_MD}
\end{figure}

The results of the case 1) are presented in Fig.~\ref{Normal_MD}. Here, $m_H = m_{H^{\pm}} = m_{A} = 200 \text{ GeV}$. $\tan \beta=1000$, $\sin (\beta-\alpha) = 0.9999$, and $m_{N_0}=20 \text{ GeV}$. Fig.~\ref{Normal_MD} clearly shows that most of the parameter space has been excluded by the $\mu \rightarrow e \gamma$ and the Planck $\displaystyle{\sum_i} m_{\nu_i}$ bounds. The deviation of the muon anomalous magnetic momenta $g-2$ cannot be explained while satifying the $l_1 \rightarrow l_2 \gamma$ bounds.

\begin{figure}
\includegraphics[width=0.45\textwidth]{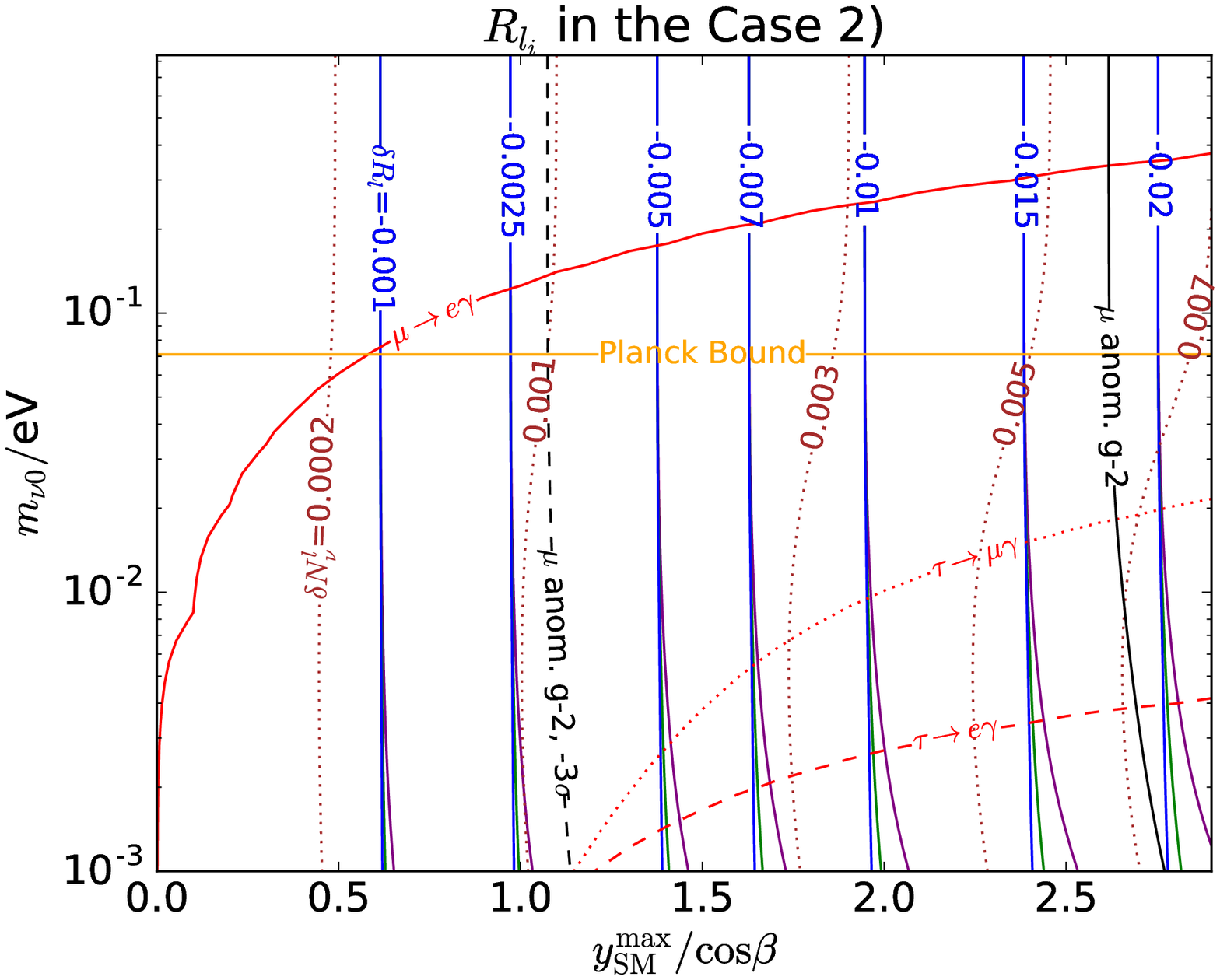}
\includegraphics[width=0.45\textwidth]{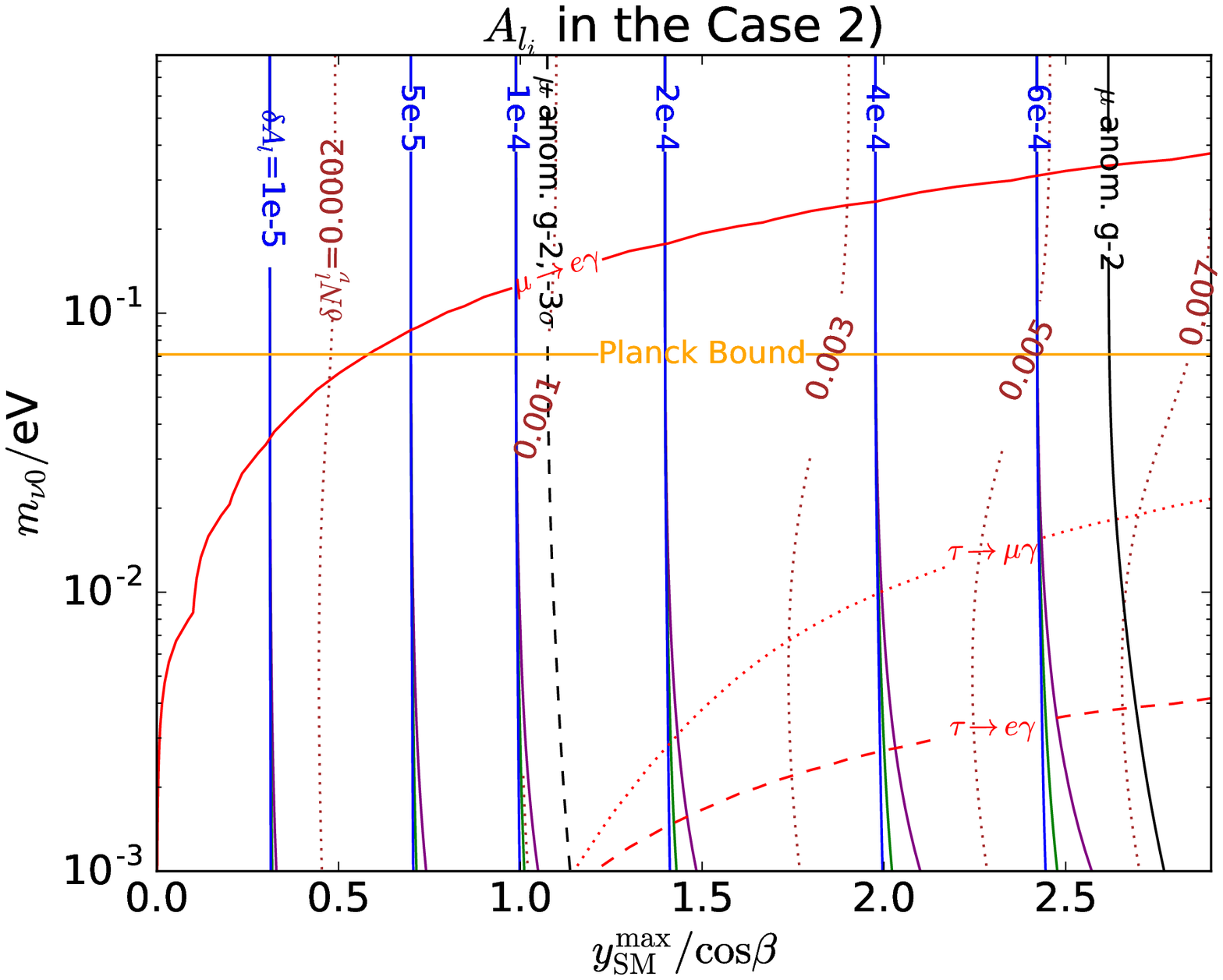}
\caption{The $R_{l_i}$ (left panel) and $A_{l_i}$ (right panel) together with the $l_1 \rightarrow l_2 \gamma$ bounds and the 3-$\sigma$ $g-2$ range. Here $m_H = m_{H^{\pm}} = m_{A} = 200 \text{ GeV}$. $\tan \beta=1000$, $\sin (\beta-\alpha) = 0.9999$, and $m_{N_0}=20 \text{ GeV}$. The purple, green, blue lines indicate the $\delta R/\delta A_{e,\mu,\tau}$ respectively.} \label{Normal_MN}
\end{figure}

The results of the case 2) are presented in Fig.~\ref{Normal_MD}. Compared with the case 1), the $e \rightarrow \mu \gamma$ bounds are somehow relaxed, however still far from explaining the deviation of the muon's anomalous magnetic momenta.

In both of the case 1) and the case 2), we can give rise to either of the $m_{N 0}$ or $m_{H^{\pm}}$ in order to suppress the branching ratio of the $l_1 \rightarrow l_2 \gamma$. However, $\delta R_{l_i}$, $\delta A_{l_i}$ and $\delta N_{\nu}$ will also be lowered, making it more difficult to be tested on the future Z-resonance experiments.

\begin{figure}
\includegraphics[width=3in]{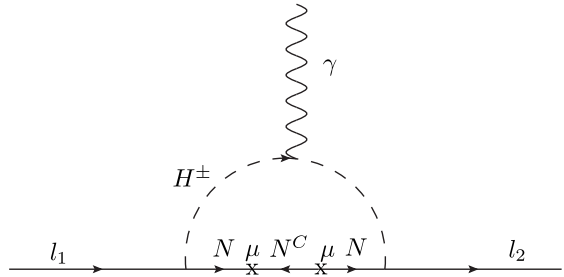}
\caption{$l_1 \rightarrow l_2 \gamma$ diagram up to the lowest order in the case 3).}
\label{l1_l2gamma_case3}
\end{figure}
As for the case 3), the $l_1 \rightarrow l_2 \gamma$ originating from the new physics sectors can be ommitted. In this case, all the leptonic FCNC effects come from the matrix $\mu$. Up to the lowest order, the diagram in Fig.~\ref{l1_l2gamma_case3} contains two insertions of the $\mu$, suppressing the $l_1 \rightarrow l_2 \gamma$ branching ratio by a factor of $\left( \frac{\mu}{m_N} \right)^4$. The complete formula is too lengthy to be presented in this paper, however in the special case when $m_N = m_{H^{\pm}}=M$, we have
\begin{eqnarray}
\sigma_{L, l_1, l_2} &=& \frac{m_{l_2} y_N \displaystyle{\sum_i} \mu_{l_1 i} \mu_{l2 i} }{1920 \pi^2 M^4}, \nonumber \\
\sigma_{R, l_1, l_2} &=& \frac{m_{l_1} y_N \displaystyle{\sum_i} \mu_{l_1 i} \mu_{l2 i} }{1920 \pi^2 M^4},
\end{eqnarray}
where $y_N$ is the diagonal element of the $Y_N \propto I$, and $l_1 \neq l_2$. Compared with the case 1) and 2), the new physics contributions to the $l_1 \rightarrow l_2 \gamma$ amplitute are too small, that we do not discuss them in the case 3).

\begin{figure}
\includegraphics[width=0.45\textwidth]{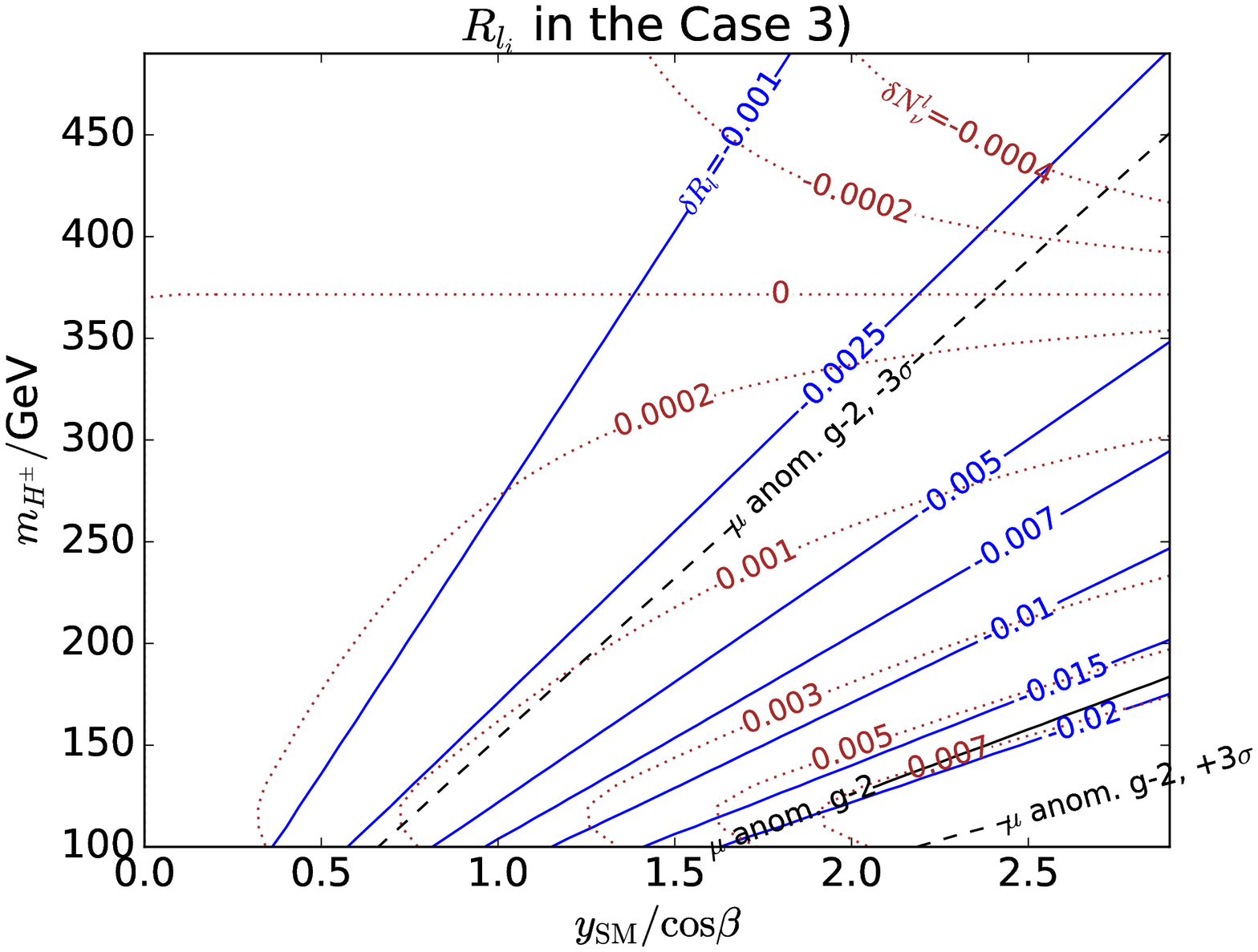}
\includegraphics[width=0.45\textwidth]{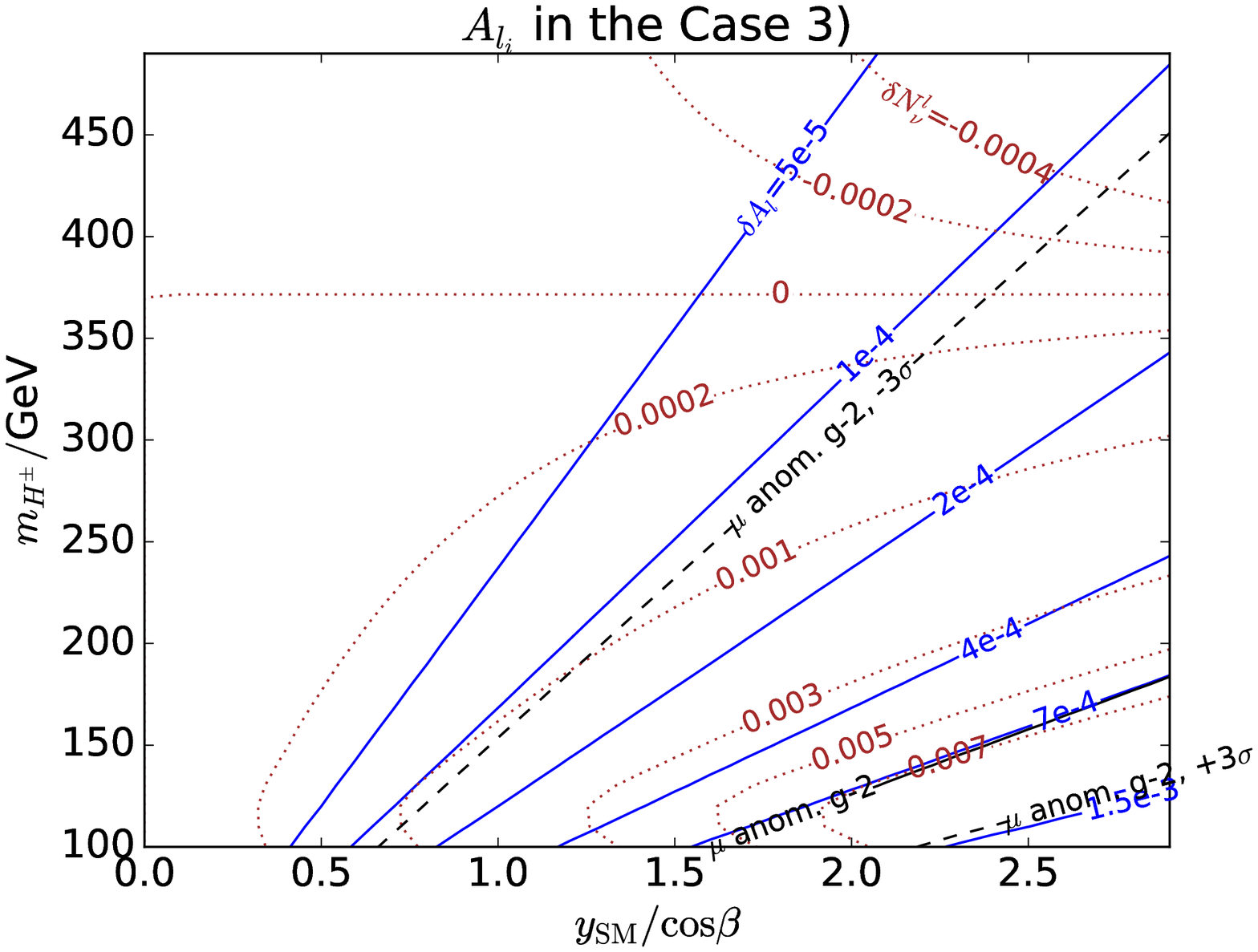}
\caption{The $R_{l_i}$ (left panel) and $A_{l_i}$ (right panel) together with the 3-$\sigma$ $g-2$ range. Here $m_H = m_{H^{\pm}} = m_{A}$ when $m_{H^{\pm}} > 125 \text{ GeV}$, however $m_H=125.1 \text{ GeV}$ and $m_{H^{\pm}}=m_{A}$ when $m_{H^{\pm}} < 125 \text{ GeV}$. $\tan \beta=1000$, $\sin (\beta-\alpha) = 0.9999$, and $m_{N_0}=20 \text{ GeV}$.} \label{mu_LightMN}
\end{figure}

\begin{figure}
\includegraphics[width=0.45\textwidth]{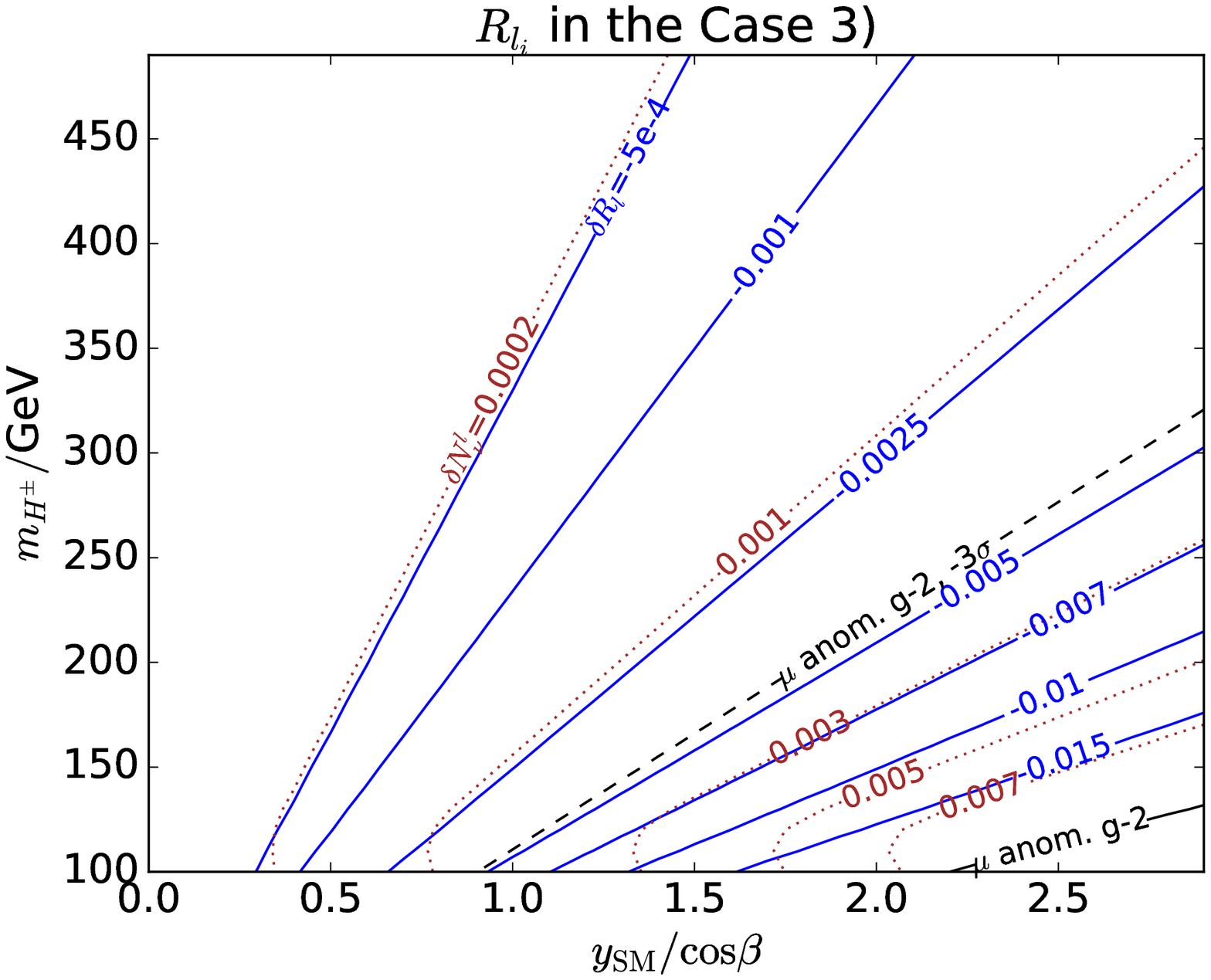}
\includegraphics[width=0.45\textwidth]{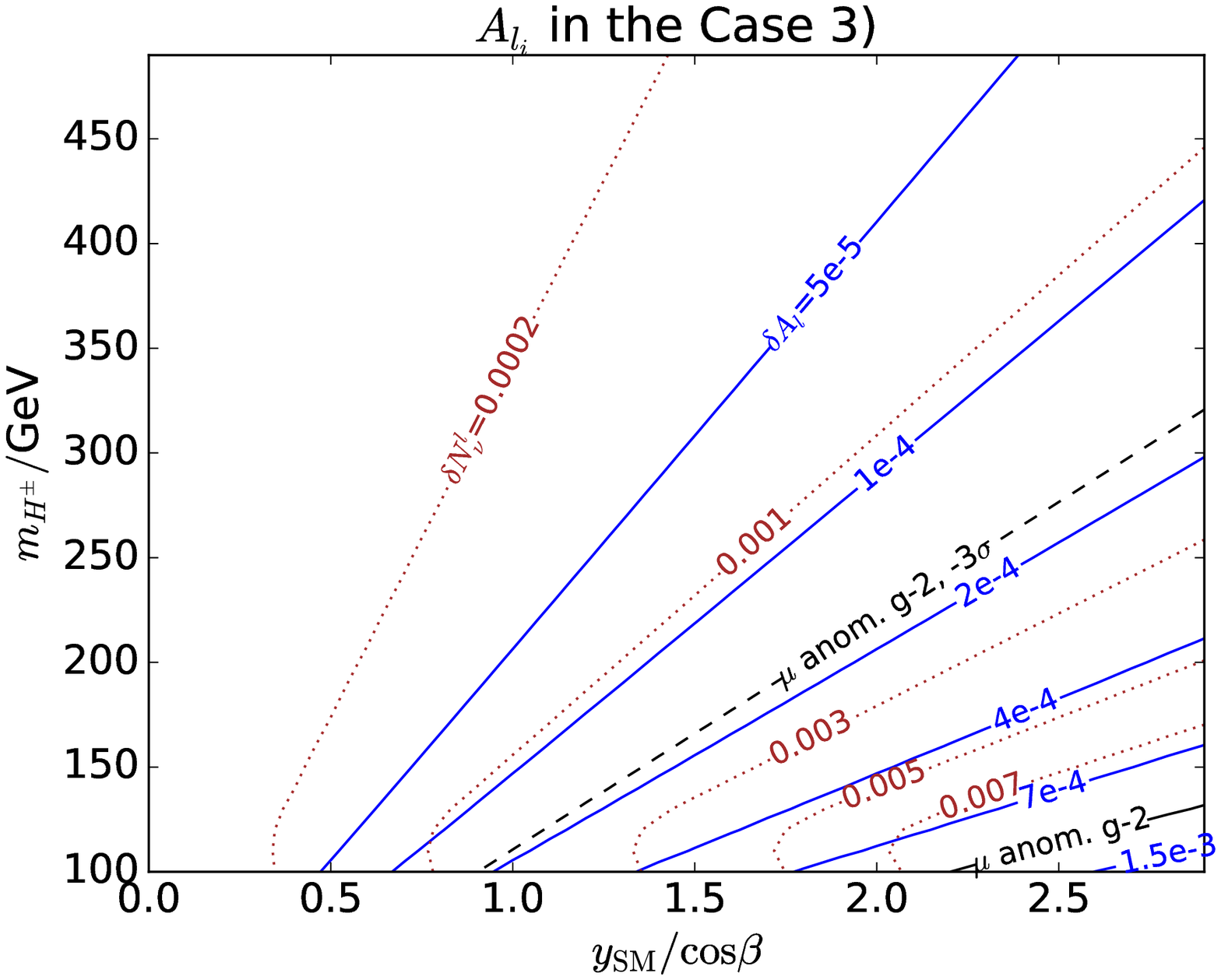}
\caption{The $R_{l_i}$ (left panel) and $A_{l_i}$ (right panel) together with the 3-$\sigma$ $g-2$ range. Here $m_H = m_{H^{\pm}} = m_{A}$ when $m_{H^{\pm}} > 125 \text{ GeV}$, however $m_H=125.1 \text{ GeV}$ and $m_{H^{\pm}}=m_{A}$ when $m_{H^{\pm}} < 125 \text{ GeV}$. $\tan \beta=1000$, $\sin (\beta-\alpha) = 0.9999$, and $m_{N_0}=m_{H^{\pm}}$.} \label{mu_MNEqualsMHpm}
\end{figure}

\begin{figure}
\includegraphics[width=0.45\textwidth]{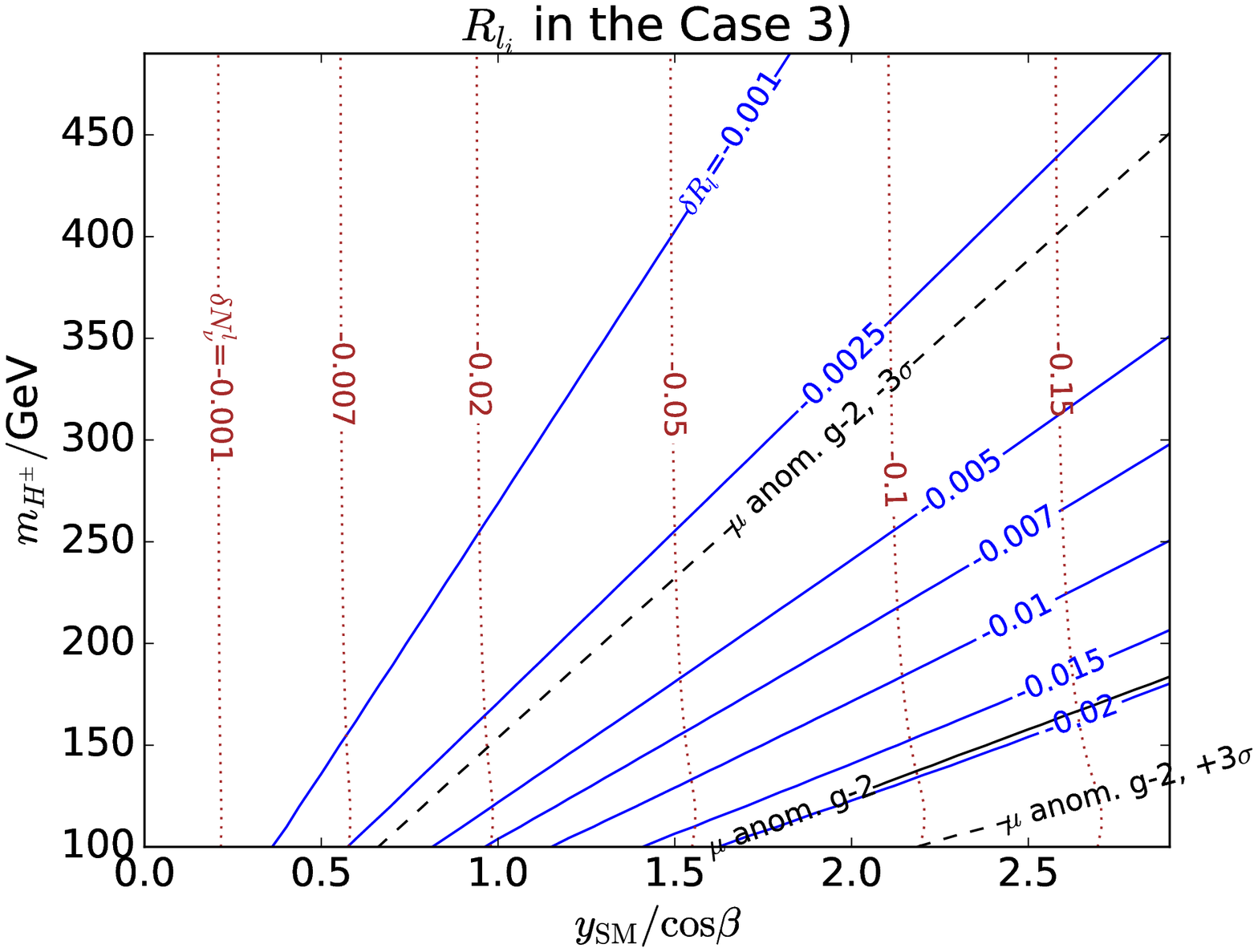}
\includegraphics[width=0.45\textwidth]{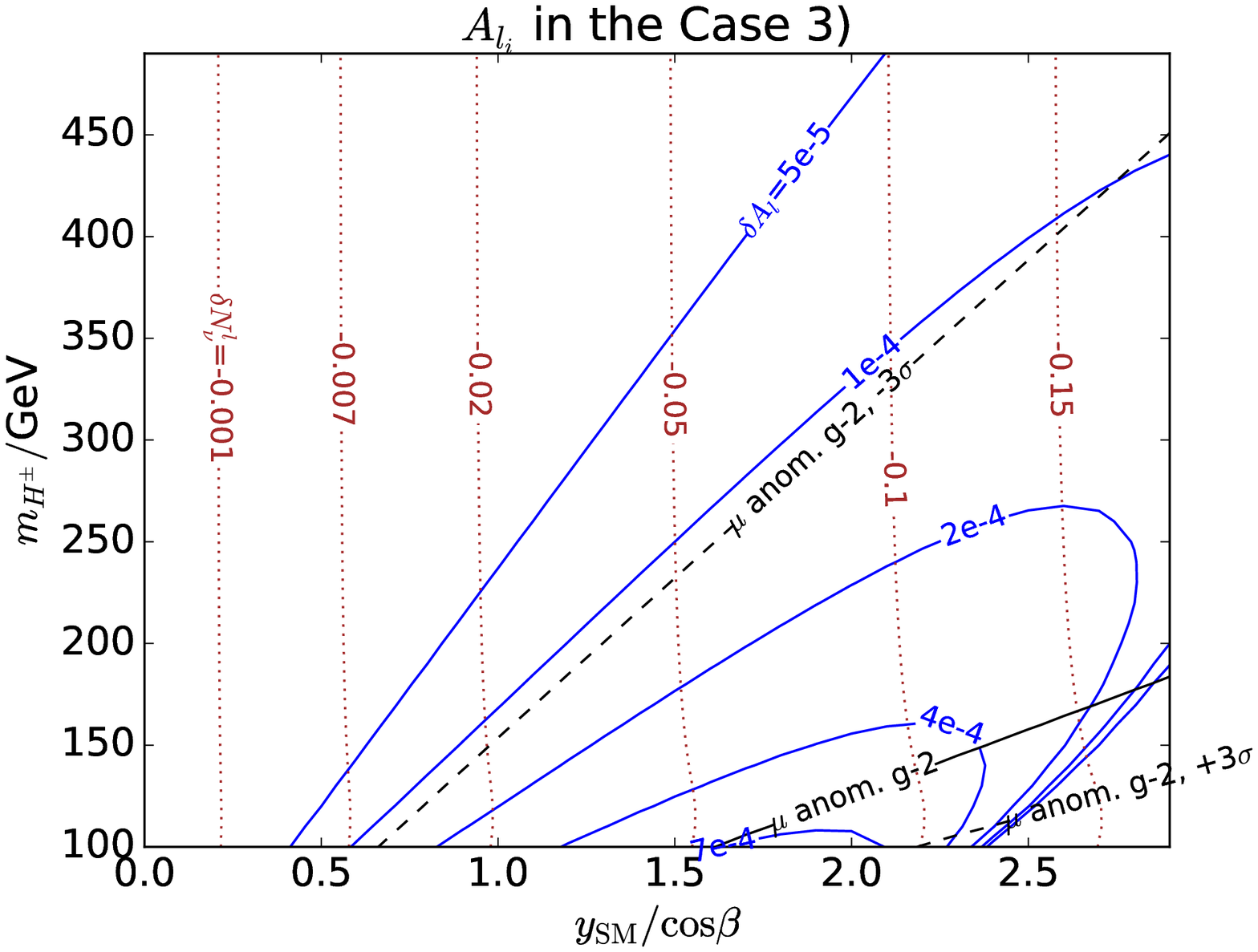}
\caption{The $R_{l_i}$ (left panel) and $A_{l_i}$ (right panel) together with the 3-$\sigma$ $g-2$ range. Here $m_H = m_{H^{\pm}} = m_{A}$ when $m_{H^{\pm}} > 125 \text{ GeV}$, however $m_H=125.1 \text{ GeV}$ and $m_{H^{\pm}}=m_{A}$ when $m_{H^{\pm}} < 125 \text{ GeV}$. $\tan \beta=300$, $\sin (\beta-\alpha) = 0.9999$, and $m_{N_0}=20 \text{ GeV}$.} \label{mu_LightMN_tb300}
\end{figure}

The results of the $\delta R_l$, $\delta A_l$ together with the 3-$\sigma$ muon's anomalous magnetic momenta range are presented in Fig.~\ref{mu_LightMN}, \ref{mu_MNEqualsMHpm} and \ref{mu_LightMN_tb300}. The model's parameter values other then the axis titles are shown in the figure captions. Notice that in the case 3), the difference between the $R_{e,\mu,\tau}$ and the $A_{e,\mu,\tau}$ are very small, so we do not distinguish them in the figures.

Compare Fig.~\ref{mu_LightMN} and Fig.~\ref{mu_MNEqualsMHpm}, it is obvious that the rise of the $m_N$ suppresses the values of the $R_l$ and the $A_l$. As for the $\delta N_{\nu}$, in most of the cases $\delta N_{\nu}>0$ because the positive one-loop contribution dominates. However when $m_N$ is small, sometimes the tree-level mixing effects between the light neutrinos and the sterile neutrinos dominate. In this case, the $\delta N_{\nu}<0$. This is more obvious when comparing the Fig.~\ref{mu_LightMN} with the Fig.~\ref{mu_LightMN_tb300}. In Fig.~\ref{mu_LightMN_tb300}, $y_{\text{SM}}$ is relatively larger due to the smaller $\tan \beta$, therefore the tree-level mixing effects always dominate so that $\delta N_{\nu}<0$. The significant difference of the $\delta A_{l}$ between the Fig.~\ref{mu_LightMN} are the Fig.~\ref{mu_LightMN_tb300} in the large $y_{\text{SM}}/\cos\beta$ area is due to the shifting of the $\theta_W$ formulated in (\ref{thetaW_shifting}), which becomes more significant when the mixings between the light neutrinos and the sterile neutrino arise.

\begin{figure}
\includegraphics[width=3in]{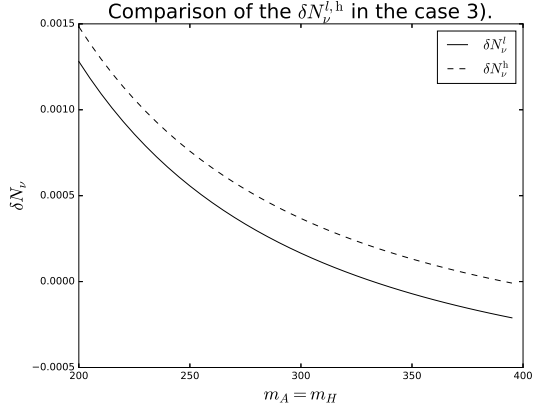}
\caption{Comparison of the $\delta N_{\nu}^{l, \text{h}}$. Here $m_H = m_{A}$ $m_{H^{\pm}} = 200 \text{ GeV}$, $\tan \beta=1000$, $\sin (\beta-\alpha) = 0.9999$, and $m_{N_0}=20 \text{ GeV}$.} \label{deltaNv}
\end{figure}

Although in the previous discussions, usually $\frac{\delta N_{\nu}^{l}}{\delta N_{\nu}^{\text{h}}} \approx 1$, this is not always the truth. Compared with the $\delta N_{\nu}^{l}$, $\delta N_{\nu}^{\text{h}}$ only receives the corrections from the neutral Higgs bosons in one-loop level. In the limit that the $m_H, m_A \rightarrow \infty$ while $m_{H^{\pm}}$ keeps small, $\delta N_{\nu}^{l}$ still receive large loop corrections due to the shifting of the $\Gamma_l$, while in this case $\delta N_{\nu}^{\text{h}}$ only receives tree-level corrections, then large deviations between $\frac{\delta N_{\nu}^{l}}{\delta N_{\nu}^{\text{h}}}$ and $1$ arise. Fig.~\ref{deltaNv} can reflect this fact in a specific area of the parameter space.

\section{Discussions}

Current experiment results show an absolute uncertainty of $\sim 0.03$-$0.05$ in the measurement of the $R_{e,\mu,\tau}$, and an absolute uncertainty of $\sim 0.005$ in the measurement of the $A_{e, \mu, \tau}$ \cite{PDG}, which is far from testing or constraining this model compared with the predicted $\delta R_{l_i}/\delta A_{l_i}$. On the future colliders, The CEPC-PreCDR \cite{CEPC_PreCDR} has mentioned that the uncertainty of the $R_{\mu}$ can be improved by a factor of roughly $\frac{1}{5}$. Both the Pre-CDR of the CEPC and ILC-GigaZ chapter in the ILC-TDR \cite{ILC_TDR_2} do not give the data for other parameters. However, it is reasonable to expect all these will be improved by roughly a of factor $\frac{1}{5}$, which can then be compared with the predicted $\delta R_{l_i}/\delta A_{l_i}$ in some of the parameter space. On the FCC-ee, Ref.~\cite{FCCee, FCCee_AFBmumu} showed that the uncertainty of $R_{l_i}$ can reach $0.001$, while the uncertainty of $A_{l_i}$ was not mentioned. However, $A_{FB}^{\mu \mu}$ can reach a relative uncertainty of 0.023\%, which can result in a similar relative uncertainty of $A_{l}$ with the assumption of $A_{e}=A_{\mu}$ and the formula $A_{FB}^{\mu \mu} = \frac{3}{4} A_e A_{\mu}$. Therefore, the performances of the $R_l$ and $A_l$ on the FCC-ee are enough to cover much of the parameter space as shown in Fig.~\ref{mu_LightMN}, \ref{mu_MNEqualsMHpm}, and \ref{mu_LightMN_tb300}. The new Z-factory proposed in Ref.~\cite{ZFactory} did not mention the measured precision of the Z-resonance parameters directly. However, compare the luminosity data given in the Ref.~\cite{ZFactory} with the Ref.~\cite{FCCee}, it is reasonable to expect a similar number of $Z$-boson can be produced in both of the two proposals. Therefore, a similar measured precision of the Z-resonance parameters can be reached.  

Another challenge is the uncertainties of the theoretical predictions of the $R_l$ and $A_l$. Currently, the theoretical uncertainty of $R_l$ is dominated by $\alpha_s$, which appears in the calculations of the $\Gamma_{\text{h}}$. In order to avoid an argument circular, we cannot use the $\alpha_s$ extracted from the Z-resonance measurements. However, In Ref.~\cite{LHEC_CDR}, LHeC itself has the potential to improve $\alpha_s$ by an order of magnitude, which will also improve the calculations of the $R_l$. As for $A_l$, the uncertainty mainly originate from the effective the weak mixing angle $\sin^2 \theta_l$. This depends on all the SM parameters, including the $\alpha$, the fine structure constant, and the Z-boson mass $m_Z$. As for the $\alpha$, if the future fittings of the uncertainty of the $\Delta \alpha^{(5)}_{\text{had}}(M_Z^2)$ (For a review about this parameter, see Ref.~\cite{PDG}. For an example calculating this from experimental data, see Ref.~\cite{a5had}) can be improved by a factor of $\frac{1}{2}$-$\frac{1}{5}$, together with all the uncertainties of other SM parameters (including $m_Z$) improved by an order of magnitude, the uncertainty of theoretical $A_l$ can also be improved and can be compared with much of the parameter space in Fig.~\ref{mu_LightMN}, \ref{mu_MNEqualsMHpm}, and \ref{mu_LightMN_tb300}.

On the future colliders, the on-shell $H^{\pm}$ might be directly produced and then decay dominantly into $l^{\pm}+N$ in this model, and $N$ then cascade decay into various SM objects that can be detected. Ref.~\cite{vTHDM_Collider2} discussed about this channel on the future HL-LHC. Their result is the $100 \text{ GeV} \lesssim m_N < m_{H^{\pm}} \lesssim 500 \text{ GeV}$ can be constrained in the future. However, heavy $m_{H^{\pm}} \gtrsim 100 \text{ GeV}$ with a rather small $m_N \ll 100 \text{ GeV}$, have not been discussed. The nearly-degenerate $m_N \approx m_{H^\pm}$ case is also difficult to be constrained. That is part of the reason why we have only presented the result when $m_N =20 \text{GeV}$ or $m_N = m_{H^\pm}$ in the section \ref{Numerical_Results}. Interestingly, we should note that when $m_N \ll m_{H^\pm}$, the sterile neutrino $N$ decay into colinear objects, which worths studying in future.

\section{Conclusions}

We proposed the $\nu$THDM with the inverse seesaw mechanisms. The Yukawa coupling involving the sterile neutrinos and the exotic Higgs bosons can take the value of order one. We have calculated the electroweak parameters $R_l$, $A_l$. The $l_1 \rightarrow l_2 \gamma$ bounds are considered, and we also calculated the predicted muon anamous momenta $g-2$. Three cases in the Tab.~\ref{Combination_Tribi} together with the flavor stuctures of the neutrinos have been considered. A large area of the parameter space in the case 1) and the case 2) are excluded by the $\mu \rightarrow e \gamma$ bound and the Planck constraint on $\displaystyle{\sum_i} m_{\nu_i}$. However, the case 3) does not receive a large correction from the new physics in FCNC parameters. By comparing the theoretical evaluations and the plans for the future collider experiments, the deviation of the $R_l$ and $A_l$ from the SM predicted values can be tested in the future collider (especially the FCC-ee) experiments. 

\begin{acknowledgements}

We would like to thank Chen Zhang, Jue Zhang, Ran Ding, Arindam Das for helpful discussions.  This work was supported in part by the Natural Science Foundation of China (Grants No.~11135003, No.~11635001 and No.~11375014), and by the China Postdoctoral Science Foundation under Grant No.~2016M600006.

\end{acknowledgements}

\newpage
\bibliography{InversevTHDM}
\end{document}